\documentclass[aps,twocolumn,superscriptaddress,floatfix,longbibliography]{revtex4-1}
\usepackage{tikz}
\usepackage{lipsum}
\usepackage{graphicx}
\usepackage[export]{adjustbox}
\usepackage{amsmath,amssymb,wasysym}
\usepackage{braket}
\usepackage{mathtools}
\usepackage{mathrsfs}
\usepackage{verbatim}
\usepackage{makecell}
\usepackage{cases}
\usepackage{tikz}

\usepackage{slashed}
\usepackage{hyperref}
\usepackage{xcite}
\usepackage{xr}

\usepackage{xcolor}
\usepackage{pgfplots}
\pgfplotsset{compat=1.18}
\makeatletter
\newcommand{\im}{\mathbf{i}}

\newcommand{\sgn}{\mathrm{sgn}}
\newcommand{\sech}{\mathrm{sech}}
\newcommand{\csch}{\mathrm{csch}}
\renewcommand{\Im}{\mathrm{Im}}

\newcommand{\BigO}{\mathcal{O}}

\newcommand{\Ehud}[1]{{\color{red} (Ehud) #1}}

\begin{document}
\title{Theory of a Strange Metal in a Quantum Superconductor to Metal Transition}

\author{Jaewon Kim}
\affiliation{Department of Physics, University of California, Berkeley, CA 94720, USA}  
\author{Erez Berg}
\affiliation{Department of Condensed Matter Physics, Weizmann Institute of Science, Rehovot 76100, Israel}
\author{Ehud Altman}
\affiliation{Department of Physics, University of California, Berkeley, CA 94720, USA}

\begin{abstract}
    Recent experiments with nanopatterned thin films revealed an unusual quantum superconductor to metal phase transition (QSMT) with a linear in temperature resistivity. In contrast, most known examples of such transitions and standard theoretical considerations predict a temperature-independent sheet resistance of order $R_Q = \frac{\hbar}{e^2}$. We propose an effective theory of a disordered superconductor which features a QSMT with robust $T$ linear resistivity at the critical point. The crucial ingredient in our model is spatial disorder in the pairing interaction. Such random pairing mirrors the emergent phase disorder seen in a recent mean field study of a microscopic $d$-wave superconductor subject to potential disorder. We also make the prediction that in such systems the diamagnetic susceptibility diverges as $\log\frac{\Lambda}{T}$, which starkly differs from the power law divergence in standard XY transitions.
    %Alternate introduction:
    %In this paper, we present a model of a disordered $d$-wave superconductor in $2+1D$ that exhibits robust linear in $T$ resistivity at a quantum superconductor to metal transition (QSMT).
    %In contrast, most known examples of such transitions and standard theoretical considerations give a constant sheet resistivity of order $R_Q = \frac{\hbar}{e^2}$.
    %The crucial ingredient in our model is spatial disorder in the pairing interaction; this random pairing interaction mirrors recent findings of emergent phase frustration in mean field studies of $d$-wave superconductors subject to conventional disorder.
    %Similar transport behaviors were reported in recent experiments with nano-patterned YBCO films.
\end{abstract}
\maketitle

\section{Introduction}

More than three decades since their discovery, the cuprates continue to pose a significant challenge to theory. In addition to high $T_c$ superconductivity, the strong electronic correlations give rise to unconventional metallic phases, including the pseudogap state in the under-doped regime \cite{Pseudogap1,Pseudogap2,Pseudogap3}, linear in $T$ resistivity near optimal doping \cite{Cuprate1,Cuprate2,Cuprate3} and a range of competing ordered states.

One way to achieve theoretical control over the complex behavior of the cuprates is to tune the system to a quantum phase transition at which the superconductivity is lost. There, one should be able to focus on a single critical fixed-point rather than an array of competing phases. But even this attempted simplification leads to intriguing surprises. In particular it was found that different ways of driving the superconductor to normal state transition result in distinct critical behaviors. A transition tuned by introducing zinc impurities, substituting for Cu atoms \cite{Fukuzumi}, follows the expected scaling of a two dimensional $XY$ model. In particular, the critical sheet resistivity approaches a constant of order resistivity quantum $R_Q \simeq \hbar/4e^2$, in the low temperature limit. This scaling is of course not special to the cuprates and is seen in many other quantum superconductor to metal transitions
\cite{Hebard1,Hebard2,Seidler,Zant,Wang}.
%This behavior is theoretically well understood as belonging to the XY universality class \cite{Fisher1,Fisher2};
%In fact, similar behaviors have been observed in most superconductor to normal state transitions in other materials \cite{Hebard1,Hebard2,Seidler,Zant,Wang};
On the other hand recent experiments with YBCO films where superconductivity is suppressed  by nano-patterning with an array of holes of varying sizes show linear in $T$ resistivity with very small residual resistivity at the critical point \cite{2022Natur.601..205Y}. %Similar behavior is seen when superconductivity is suppressed by a magnetic field in the overdoped regime \cite{2009Sci...323..603C,2012PNAS..109.8440B}.
%\Ehud{Is this really true? What I see in Fig. 2E is at about the resistivity quantum and not clearly $T$ linear. Perhaps we can instead refer to the field tuned transition near optimal doping?}.
%and magnetic fields \cite{2012PNAS..109.8440B}.
Previous theoretical work on disordered $d$-wave superconductors could capture a small residual critical resistivity $R_0\ll R_Q$, but not the linear temperature dependence\cite{Herbut,Podolsky}. 

\begin{figure}
    \centering
    \includegraphics[width=0.9\columnwidth]{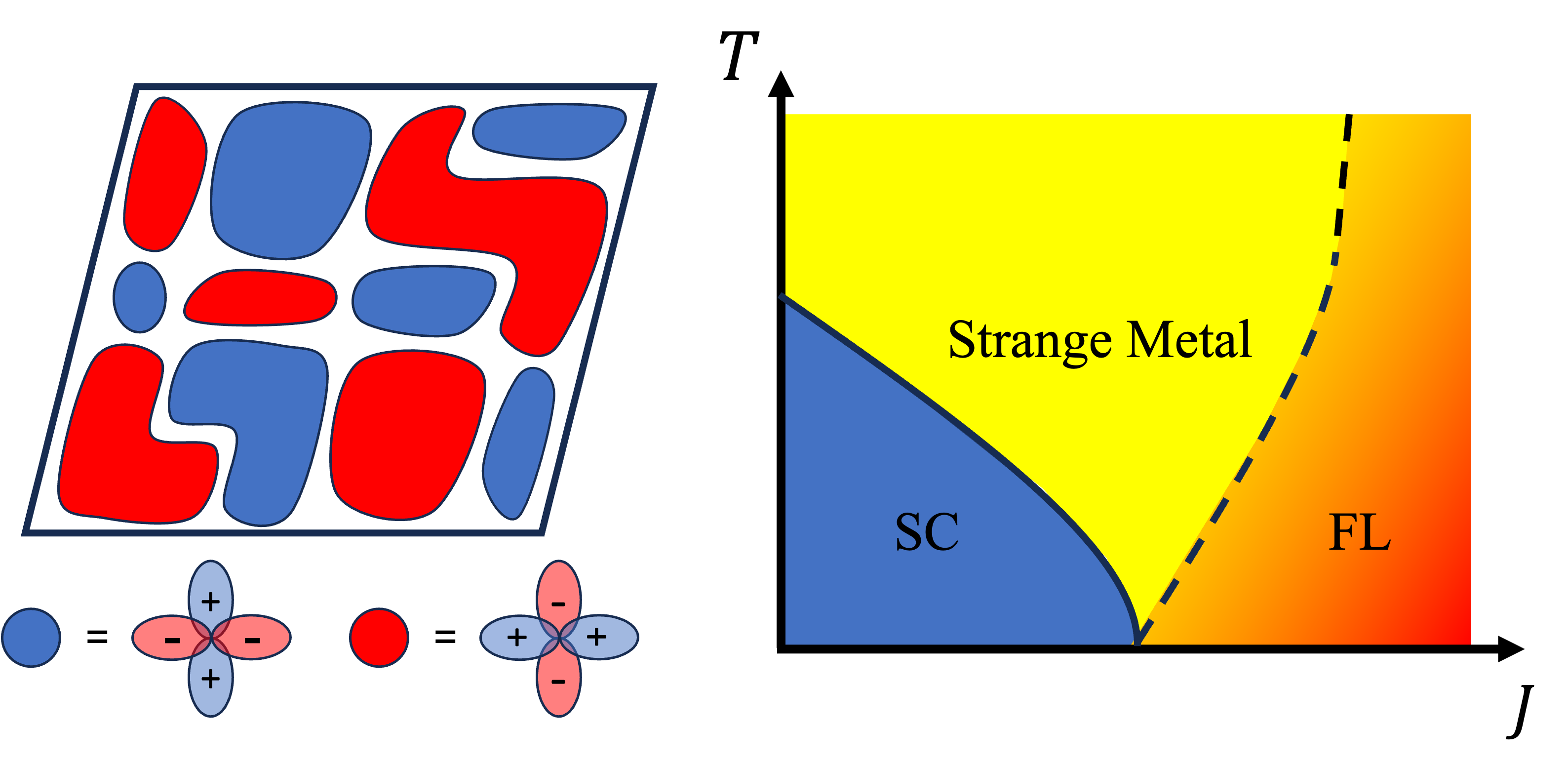}
    \caption{(\textit{Left}) An illustration of our model. Our model is a disordered $d$-wave superconductor with spatially random sign changes of the order parameter. (\textit{Right}) Schematic phase diagram of our model. The strange metal behavior occur at the critical point of the quantum superconductor to metal transition.}
    \label{fig:mainpic}
\end{figure}

In this paper we present a model of a disordered $d$-wave superconductor that leads to $T$ linear resistivity at the critical point over a broad temperature range. As in the earlier theories the effective model we study includes a fluctuating $d$-wave order parameter coupled to gapless fermions occupying a Fermi sea. The key new element is that we take the dominant disorder driving the transition to be a random pairing interaction, or more precisely disorder in the strength and sign of the coupling between the fermions and the fluctuating $d$-wave pairing field (see Fig. \ref{fig:mainpic}). 

Including this kind of disorder is motivated by recent mean field studies of a d-wave superconductor, showing that short range correlated potential disorder gives rise to emergent longer scale disorder in the sign of the pairing field \cite{dunghai}. To account for the longer scale of this disorder we take it to be correlated over a length $\xi > 2\pi/k_F$.
%Crucially, such disorder cuts off the BCS instability and thus it allows the QSMT to take place even in the extreme case where only this disorder is included (i.e. without disorder in the single particle terms). 

We solve the effective model in a controlled large $N$ limit, where the numbers of fermion and boson (order parameter) species are taken to infinity together. The main advantage of this approach, developed in Refs. \cite{Kim:2020jpz,Aldape:2020enq,esterlis2021large,patel2} to describe critical points in itinerant fermion systems, is that it preserves the mutual feedback between the critical fluctuations (order parameter fields) and the fermions. This allows us to capture a strong coupling critical point, exhibiting linear in $T$ resistivity within the large $N$ saddle point. 

In addition to the $T$ linear resistivity our model also shows linear magnetoresistance at the critical point with a universal relaxation rate $1/\tau \sim \mu_B B$. Such behavior has been observed in a number of correlated materials \cite{LSCO,2009Sci...323..603C,Hayes,2022NatPh..18.1420A,Jaoui} and in the patterned cuprate films discussed above.
Within our model, however, the linear magnetoresistance is found only when the cyclotron frequency $\omega_{cf}$ is in the rather restricted range  $\omega_{cf} \in k_B T[(k_F\xi)^{-4},(k_F\xi)^{-3}]$, which may be too small to explain the experiments.

The rest of the paper is organized as follows. In section \ref{sec:model} we introduce the model and give an overview of the main results. In section \ref{sec:dsol} we detail the solution for the single particle Greens functions in the large $N$ limit. This allows us to derive the conductivity in section \ref{sec:dtransport} and the magnetic susceptibility in section \ref{sec:DS}. We conclude with a discussion in section \ref{sec:discussion}.

\section{Model and overview}
\label{sec:model}
In this section we present and motivate the effective model we employ to understand the superconductor to metal quantum phase transition in patterned cuprate films and summarize the main results from its analysis.  
The model consists of a field theory for a complex field describing $d$-wave pair fluctuations, coupled to a Fermi surface through a random pairing interaction in 2+1 dimensions.
The action is given by,
\begin{subequations}
\label{eq:dL}
\begin{align}
    &\mathcal{S} = S_b + S_f + S_{int}\,, \textrm{ where } \nonumber \\
    &S_b = \int_\tau \sum_x \Delta_b^0 |\Psi_x|^2 + \frac{U}{2} |\Psi_x|^4 + \sum_{\braket{x x'}} t_b \Psi_x^\dagger \Psi_{x'}
    \label{eq:dLb} \\
    &S_f = \int_{\tau,k} c_k^\dagger \left(\partial_\tau + \epsilon_k - \mu \right)c_k
    \label{eq:dLf} \\
    &S_{int} = \int_\tau \sum_x J g_x \Psi_{x} \left(c^T_{x} \sigma^y c_{x+a\hat{x}} - c^T_{x} \sigma^y c_{x+a\hat{y}} \right) + c.c.
    \label{eq:dLint}
\end{align}
\end{subequations}
Here, $\Psi_x$ is a complex boson field at position $x$ describing the Cooper pair fluctuations, and $c_x^T = (c_{x,\uparrow},c_{x,\downarrow})$, the fermion field at position $x$.
For simplicity, in \eqref{eq:dLf} we consider a quadratic dispersion $\epsilon_k = k^2/2m_f$.
$g_x$ is a quenched disorder field with zero mean and correlation length $\xi$ that satisfies
\begin{equation*}
    \overline{g_x g_{x'}} \simeq e^{-|x-x'|^2/4\xi^2}
\end{equation*}

The Ginsburg-Landau action $S_b$, in itself, is a standard description of a superconductor to insulator quantum phase transition.
The disordered coupling of the bosonic field to a Fermi surface is designed to stabilize a metallic normal state in place of the insulator. Note that a uniform coupling to the fermions cannot achieve this. Even in the gapped phase of $S_b$, integrating out $\Psi$ in presence of such uniform coupling generates attractive interactions between the fermions leading to a BCS instability. Hence the transition is eliminated and the system is superconducting throughout. 
The random coupling $g_x$ provides an IR cutoff to the BCS instability at the length scale $\xi$.
As a result, superconductivity onsets only beyond a finite coupling strength $J$.
The quantum phase transition can also be tuned at constant $J$ by varying the disorder correlation length $\xi$.

It is worth noting that the conventional disorder present in physical systems at short scales couples directly to the fermion density rather than to the pairing field. The pairing disorder in our model is a phenomenological description of inhomogeneity that can emerge at longer scales due to the interplay of the short-range disorder and d-wave pairing.
Self consistent mean field calculations have shown that this interplay gives rise to geometrically frustrated phase disorder at longer scales, evidenced by emergence of isolated $\pi$ vortices \cite{dunghai}.
Instead of this geometric frustration, the random sign of the coupling in the model \eqref{eq:dL} induces internal frustration between the boson and fermion species. 
%In Section \ref{subsec:ddd} we study the effect of adding weak diagonal disorder to the model \eqref{eq:dL}.

To facilitate a controlled solution we consider an extension of the model to a large number $N$ of fermion and boson species. Accordingly, the Yukawa coupling $g$ between the species is promoted to a random tensor. Similar techniques have been developed recently to deal with strongly coupled quantum critical points, first in Dirac systems \cite{Kim:2020jpz} and later in quantum critical points with a metallic Fermi surface \cite{Aldape:2020enq,esterlis2021large,patel2}.
Keeping the ratio of fermionic and bosonic species constant in the large $N$ limit ensures that the mutual feedback between them is preserved. In particular,
the bosons are Landau damped due to coupling to the fermions, generating a self-energy that scales as $\alpha|\omega|$ at the QCP.
At the same time the fermions become a marginal Fermi liquid, due to coupling to the critical bosons, exhibiting a self-energy that scales as $\Sigma(k,\omega)\sim\gamma_k\omega \log\omega$. 

The fermion self-energy implies a scattering rate that grows linearly with temperature as  $1/\tau_k \approx \gamma_k T$.
The angle dependence of $\gamma_k$ stems from the $d$-wave nature of the coupling and for sufficiently smooth disorder (i.e. $k_F\xi\gg 1)$ leads to strongly suppressed scattering at the nodes $1/\tau_{\textrm{node}} \sim T/(k_F\xi)^{3}$.
As a result, the nodes become a significant bottleneck in momentum relaxation, leading to the inverse transport time $\frac1{\tau_{\textrm{tr}}} \sim (k_F \xi)^{-4} T$.
Thus, the quasiparticle scattering rate is linear in $T$, but sub-Planckian for smooth disorder.

In section \ref{sec:dtransport} we calculate the conductivity using the quantum Boltzmann equation.
The conductivity is dominated by the quasiparticle (fermion) current and is given by $\sigma \approx (k_F \xi)^4 e^2 \epsilon_F/T$;
%Due to a bottleneck in relaxation at the nodes, their momentum relaxation rate scale as $(k_F \xi)^{-4} T$, making them long lived.
The critical pairing fluctuations, on the other hand, contribute a sub-leading logarithmic temperature dependence $\sigma_b\sim \log (\Lambda/T)$.
%The two conductivities add  %parallel; ergo, at low $T$ the total %transport is dominated by the %conductance from unpaired fermions.
%Consequently, the total resistivity scale as 
Hence the system exhibits strange metal behavior with $\rho \simeq (k_F \xi)^{-4} (T/e^2\epsilon_F)$.

In section \ref{subsec:ddd} we investigate the effect of adding weak conventional potential disorder coupled to the local fermion density at short scales.   
This scattering generates a small constant decay rate $1/\tau_0$ that ultimately leads to a residual resistivity $\rho_0 \sim (e^2 \epsilon_F \tau_0)^{-1}$ at zero temperature.
Scattering from potential disorder also eliminates the nodal bottleneck for quasi-particle relaxation, which in turn alters the slope of the $T$ linear resistivity below a crossover temperature scale.
%to $k_F\xi^{-3}/(e^2 \epsilon_F)$ for temperatures below $T_0= (k_F\xi)^4/\tau_0$. At higher temperature the slope returns to  $\frac{(k_F\xi)^{-4}}{e^2 \epsilon_F}$ as in absence of the potential disorder.
%Interestingly, the linear $T$ dependence to zero temperature gives $\rho_*\sim (k_F\xi)^3/\tau_0$, which over-estimates the true residual resistivity by a factor of $(k_F\xi)^3$. 
%[What is the saturation value?] Interestingly, the true saturation value  saturates to is order $(k_F\xi)^{-1}$ smaller than the value extrapolated from the high temperature $T$ linear regime. This is due to the fact that the main contribution to charge transport comes within small regions of the fermi-surface near the nodes.
%For similar reasons, at intermediate temperature scale in the crossover from linear in $T$ and saturation, the resistivity scales as $\sqrt{T}$.

In section \ref{subsec:MR} we study the magnetoresistance at the critical point. 
%We observe diverse scaling behavior as a function of the ratio between the fermion cyclotron frequency and temperature, $b = \omega_{cf}/T$.
In particular, we find a linear in $B$ magnetoresistance for the range of magnetic fields such that the fermion cyclotron frequency $\omega_{cf}$ is within the range $(k_F\xi)^{-4}T \ll \omega_{cf} \ll (k_F \xi)^{-3} T$.
%In this regime, the nodal bottleneck in relaxation is overcome through the fermion cyclotron motion;
%The resulting relaxation rate is linear in $B$ with a universal coefficient given by the Bohr magneton $\frac{e}{m_f}$, which in turn gives rise to linear magnetoresistance.

In section \ref{sec:DS} we study the diamagnetic susceptibility $\chi(T)$ in the critical regime. We predict a logarithmic divergence $\chi\sim \log(\Lambda/T)$ which is contrasted with the power law divergence
expected in a conventional XY quantum critical point in $2+1D$.
This much slower divergence is a result of  strong Landau damping of the order parameter fluctuations and offers a sharp testable prediction of our theory.

\section{Large N Model and Solution}
\label{sec:dsol}
%It is a hard task to compute the conductivity of \eqref{eq:dL} in the strongly interacting regime.
In this section we describe the calculation of the fermion and boson (pairing field) Green's functions in the special large $N$ limit described above. To this end we extend the model to $N$ flavors of (spinful) fermions  $c_i, i=1,..., N$ and the same number N of boson flavors $\Psi_n, n = 1,..., N$ \footnote{In principle, the number of bosons and fermions can be different as long as they are taken to infinity together (i.e. with a constant ratio), but this detail does not greatly affect the physics of our model.}.
Accordingly, the disorder in the pairing $g$ is also promoted to a random rank 3 tensor $g_{ij}^n$.
In this large $N$ limit, we may solve a closed set of saddle point (Schwinger-Dyson) equations to obtain the exact boson and fermion correlation functions and their corresponding self-energies.
Since we keep the ratio of the number of fermion to boson flvaors constant, the correlation functions and the self-energies that we compute from the saddle point equations contain the mutual feedback between the bosons and fermions.
%In turn, the correlation functions that we find in this section will be used in the next section to calculate the transport characteristics using linear response theory.

Upon generalizing \eqref{eq:dL} to $N$ flavors of fermions and bosons, our action is given by,
\begin{widetext}
\begin{equation}
\begin{split}
    \mathcal{S}_N (c^\dagger,c,\Psi) &= \int_k \sum_{i=1}^{N} {c_{ik}^\dagger}(\partial_\tau + \epsilon_k-\mu)  c_{ik} + \sum_{n=1}^{N}\left(\Delta_b^0 + \frac{k^2}{2m_b}\right) \Psi_{nk}^\dagger \Psi_{nk} + \int_{x} \frac{U}{2N} \left( \sum_{n=1}^N \Psi_{nx}^\dagger \Psi_{nx} \right)^2 \\
    & + \int_{q,q',k} \sum_{ij,n}^N J_k g_{ij,q}^n \Psi_{q'}^n  \left({c}^T_{i \frac{q+q'}{2}+k} \sigma_y c_{j \frac{q+q'}{2}-k} \right)_x + c.c. \,, \textrm{ where } J_k = J(\cos k_x a - \cos k_y a) 
    \label{eq:largeNmodeld}
\end{split}
\end{equation}
\end{widetext}
Here, we have written the pairing interaction in momentum space.
As such, the pairing interaction has a momentum dependence $J_k$, which vanishes at the nodes of $k_x = \pm k_y$.
In addition, the disorder in momentum space $g_q$ is of zero mean and its variance satisfies,
%\footnote{The precise relation is $\overline{g_{ij,q}^n g_{kl,q'}^m} = \frac{\pi \xi^2}{N^2} \delta_{nm} \delta_{il} \delta_{jk} e^{-\xi^2 q^2}$, but we are approximating the gaussian distribution as a Heaviside function for simplicity in computation.},
\begin{equation}
\begin{split}
    \overline{g_{ij,q}^n g_{j'i',q'}^{n'}} &= \frac{\pi \xi^2}{N^2} \delta_{nn'} \delta_{ii'} \delta_{jj'} \delta(q+q') e^{-\xi^2 q^2} \,, \\
    %&\simeq \frac{\xi^2}{N^2} \delta_{nm} \delta_{il} \delta_{jk} \delta(q+q') \Theta(\pi/\xi-|q|)
    \label{eq:ddisorderp}
\end{split}
\end{equation}
Accordingly, the disorder in the pairing $g$ can scatter momentum up to $\BigO(\pi/\xi)$.
%In what follows, we shall treat the gaussian momentum dependence of \eqref{eq:ddisorderp} to a Heaviside step function with cutoff $\pi/\xi$ for simplicity.

\begin{comment}
As we will show, the bosons are strongly Landau damped due to interactions with the fermions.
Furthermore, the interaction also renormalizes the boson mass to be extremely light.
%this constrains the boson low energy degrees of freedom to be constrained to small momenta $q \ll \frac1\xi$.
In turn, interactions with the critical bosons cause the fermions to become a marginal Fermi liquid at the QCP; they obtain a self-energy of $\gamma_k \omega \log \frac{\Lambda}{\omega}$.
The $d$-wave nature of the pairing leads to the angle dependent prefactor $\gamma_k$, which generally scales as $\frac{1}{k_F\xi}\cos^2 2\theta_k$.
$\gamma_k$ is smallest at the nodes, where it scales as $(k_F\xi)^{-3}$.
This strong suppression of the self-energy at the nodes lead to nodal quasi-particles dominating charge transport.
\end{comment}

\begin{figure}
    \centering
    \includegraphics[width = 0.8\columnwidth]{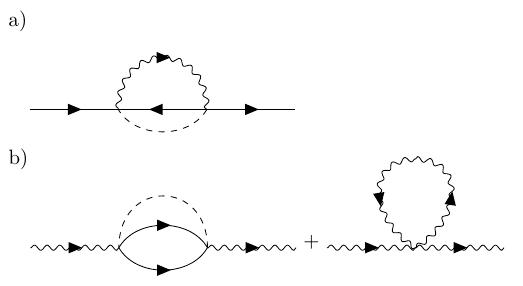}
    \caption{Graphic representation of the large $N$ saddle point equations. Solid (wavy) lines denote fermions (bosons, resp.). Dashed lines denote disorder contraction. a) Fermion self-energy diagram. b) Boson self-energy diagrams.}
    \label{fig:SD}
\end{figure}

We now perform disorder averaging and take the saddle point, which gives the following Schwinger-Dyson equations (Also illustrated pictorially in Fig.\ref{fig:SD}).
%Solving it self-consistently, we can find the exact boson and fermion correlation functions and their corresponding self-energies while preserving mutual feedback.

\begin{widetext}
\begin{equation}
\begin{split}
    & G(k,\omega) = \frac{1}{\im\omega + \Sigma(k,\omega) - (\epsilon_k-\mu)} \,, \ \ \Sigma(k,t) = \int_{|q'| \lesssim \frac\pi\xi} \int_{q}  2 J^2_{k-\frac{q+q'}{2}} \xi^2 F(q,t) G(-k+q+q',-t) \,, \\
    & F(q,\omega) = \frac{1}{\Delta_b^0-\Pi_f(q,\omega)+q^2/2m_b} \,, \ \ \Pi = \Pi_b + \Pi_f \,, \textrm{ where } \\
    &\Pi_f(q,t) = \int_{|q'| \lesssim \frac\pi\xi} \int_{k} 2J^2_{k-\frac{q+q'}{2}} \xi^2 G(k,t) G(-k+q+q',t) \,, \ \ \Pi_b(q,\omega) = -U \int_{k,\omega}F(k,\omega) \,.
\end{split}
\label{eq:dSDeq}
\end{equation}    
\end{widetext}
Here, we denote with $G = \frac{1}{N} \sum_i \braket{c_i c^\dagger_i}$, the fermion Green's functions, and $F = \frac{1}{N} \sum_n \braket{\Psi_n \Psi^\dagger_n}$, the boson Green's functions.
$\Sigma$ and $\Pi$ denote the fermion and boson self-energies, respectively.
We work from the metallic side of the transition, so that superconducting correlation functions such as $\braket{\psi \psi}$ are zero.

The fermion self-energy $\Sigma$ is simply a bubble diagram of a fermion and boson propagator.
On the other hand, the boson self-energy $\Pi$ is a sum of two parts, $\Pi_b$ and $\Pi_f$.
The boson contribution $\Pi_b$ is due to the quartic boson interaction and is a tadpole diagram.
Conversely, the fermion contribution $\Pi_f$ is due to the random pairing interaction and is a bubble diagram of two fermion propagators.

It is important to note that due to \eqref{eq:ddisorderp}, the disorder lines can only carry momentum $q' \lesssim \frac{\pi}{\xi}$.
This fact is reflected in the integral over $q'$ in \eqref{eq:dSDeq} for $\Sigma$ and $\Pi_f$.

%It is worth nothing that the quartic boson interaction $U$ is renormalized.
%However, the renormalization flow is logarithmically slow, and in this work, we shall essentially treat it as a constant.
%it has a strong momentum and frequency dependence that leads to renormalization of the boson mass, dynamics, etc.
%Solving \eqref{eq:dSDeq} self-consistently, we can find the boson and fermion correlation functions and their corresponding self-energies while preserving mutual feedback.
%In turn, these correlation functions and the self-energies will be used in the next section to calculate the conductivity.

We begin solving \eqref{eq:dSDeq} by first determining the boson self-energy $\Pi$.
In particular, let us first focus on its frequency and momentum dependence.
This comes solely from $\Pi_f$, the self-energy due to the random pairing interaction;
%As we will show, the originally non-dynamical bosons become dynamic and their mass becomes extremely light due to interactions with the fermions.
%This is because the frequency and momentum dependence of the boson self-energy comes from $\Pi_f$, the self-energy due to the random pairing interaction;
since $\Pi_b$ is a tadpole diagram, it is a constant independent of frequency and momentum.

For small $q$ and $\omega$, $\Pi_f(q,\omega)$ is given as (See Appendix.\ref{app:SD} for details),
\begin{equation}
\begin{split}
    &\Pi_f(q,\omega) = \Pi_f(0) - \frac{q^2}{2M} - \alpha |\omega| + \im\eta \omega \,, \textrm{ where } \\
    & \frac{1}{M} = \frac{J^2 \xi^2 k_F}{8\pi^2 v_F} \,, \ \alpha = \frac{J^2 \xi k_F}{4\pi v_F^2}\,, \ \eta \simeq \frac{J^2 \log \frac{\Lambda_f}{v_F/\xi}}{v_F^2} \,.
    \label{eq:bosonPid}
\end{split}
\end{equation}

\eqref{eq:bosonPid} has two important consequences. First, the boson propagator which was originally independent of frequency now obtains a frequency dependence; it is Landau damped ($\alpha$) and particle-hole asymmetric ($\eta$).
In particular, the Landau damping is very strong compared to the particle-hole asymmetry with $\frac{\alpha}{\eta} \simeq \frac{k_F \xi}{\log k_F \xi} \gg 1$.

Second, the momentum dependence of $\Pi$ means that the boson mass $m_b$ is renormalized to $M_b = (m_b^{-1} + M^{-1})^{-1}$.
Together with the particle-hole asymmetry, the effective boson mass $\tilde M_b$ is given by $\eta M_b \simeq \frac{\log k_F \xi}{(k_F \xi)^2} m_f$.
For $k_F\xi \gg 1$, $\tilde M_b$ is extremely light;
this means that the boson low energy degrees of freedom are essentially restricted to small momenta of order $\BigO(1/\xi)$.

We now turn to $\Pi(0)$, the boson self-energy at zero frequency and momentum.
It renormalizes the boson gap $\Delta_b$ as,
\begin{subequations}
\label{eq:bosongapd}
\begin{align}
    & \Delta_b = \Delta_b^0 - \Pi_b(0) - \Pi_f(0) \,, \textrm{ where } \\
    & \Pi_b(0) = -U \int_{k,\omega} F(k,\omega) \,,
    \label{eq:Pib0d} \\
    & \Pi_f(0) = \int_{q' \lesssim \pi/\xi} \int_{k,\omega} 2J_{k-\frac{q'}{2}}^2 G_{k+q',\omega} G_{-k,-\omega} \nonumber \\
    & \quad\quad \ \ \simeq \frac{J^2 k_F}{v_F} \log k_F \xi
    \label{eq:Pif0d}
\end{align}
\end{subequations}
Here, both $\Pi_b$ and $\Pi_f$ participate in the boson gap renormalization.
The boson contribution $\Pi_b(0)$ \eqref{eq:Pib0d} increases the gap.
On the other hand, the fermion contribution $\Pi_f(0)$ \eqref{eq:Pif0d} decreases the gap and drives the bosons toward superconductivity.
%Note that $\Pi_b$ computed from \eqref{eq:Fqcp} saturates to $\sim \frac{UM_b}{\alpha} \Lambda$, where $\Lambda$ is the boson UV cutoff.

Crucially, $\Pi_f$ is IR convergent.
This is in contrast to an IR divergence seen in standard BCS theories, where the ground state inevitably becomes superconducting due to the divergence.
Within our model, the disorder in the pairing $g$ is providing an IR cutoff lengthscale $\xi$ that eliminates this BCS instability.
Therefore, we have a quantum phase transition between the superconducting phase ($\Delta_b < 0$) and the metallic phase ($\Delta_b > 0$) as we tune the interaction strength $J$ or the disorder correlation length $\xi$. In the rest of this paper we focus on the critical point, defined by $\Delta_b=0$.

\begin{comment}
%Furthermore, we will take the bare boson gap $\Delta_b^0 \sim \epsilon_F$, and bare boson mass $m_b \sim m_f$.
%This condition implies that the bosons before renormalization are short ranged.
%Such an assumption is appropriate, for it ensures that the four fermion interaction upon integrating out the bosons is also short ranged.

%Let us make a rough estimate for the critical value of $J$. At the critical point, $\Delta_b = \Delta_b^0 - \Pi_f(0) - \Pi_b(0)$ vanishes by definition. For small boson quartic interactions, the boson contribution $\Pi_b$ is small, so $\Delta_b^0 \simeq \Pi_f(0)$. Since the bare gap $\Delta_b^0 \sim v_F k_F$, we conclude that at criticality $J \sim v_F/ (\log k_F \xi )^{\frac12}$.
\end{comment}

Putting the above results together gives us the boson propagator at the QCP,
\begin{equation}
\begin{split}
    & F(k,\omega) = \frac{1}{k^2/2M_b + \alpha |\omega| -\im\eta\omega}\,.
    \label{eq:Fqcp}
\end{split}
\end{equation}

We can now use the boson propagator to determine the fermion self-energy $\Sigma$,
which is given by the convolution of the boson and fermion propagators as shown in (Fig.\ref{fig:SD}a).
At the QCP, this convolution is dominated by the critical bosons \eqref{eq:Fqcp}, leading to a marginal Fermi liquid \cite{Chandra} self-energy, albeit with a momentum dependent prefactor $\gamma_k$ (See Appendix.\ref{app:SD}):
\begin{equation}
\begin{split}
    &\Sigma(k,\omega) \sim \im\gamma_k \omega\log \frac{\Lambda}{|\omega|} \,, \\
    & \textrm{ where } \gamma_k \simeq  \frac{2}{\pi} \left(\frac{2\pi}{k_F \xi} + \cos 4\theta_k \sin \frac{2\pi}{k_F \xi} \right)\,.
    \label{eq:SigmaE}
\end{split}
\end{equation}
$\gamma_k$ generally scales as $\frac{\cos^2 2\theta_k}{k_F \xi}$, due to the $d$-wave nature of the interaction that has a momentum dependence of $J_k \sim J \cos 2\theta_k$.

Importantly, even at the nodes of $J_k$ at $\theta_k = \pm \frac{\pi}{4}, \pm\frac{3\pi}{4}$, $\gamma_k$ is nonzero, but rather takes the small value $\gamma_k \sim (k_F \xi)^{-3} \ll 1$.
The nonzero scatterring rate at the nodes is due to the momentum carried by the pairing disorder $g$ that allows for scattering within angles $(k_F\xi)^{-1}$ of the nodes to contribute to the nodal self-energy.

%This strong suppression of the self-energy at the nodes means that the nodal degrees of freedom are longest lived and dominate the transport.

Having derived the critical propagators, we now generalize the results to non vanishing temperatures in the critical regime.
%We do this through analytical continuation from the QCP and working in the Keldysh formalism.
%Doing this allows us to compute the quasi-particle relaxation rate $\tau_k^{-1}$ as a function of temperature.
%As we will show, $\tau_k^{-1}$ is sub-Planckian when the disorder correlation length $\xi$ is large; as $\xi$ gets shorter, the relaxation rate becomes faster until it saturates to Planckian dissipation when $k_F\xi \sim \BigO(1)$.
%We do this to obtain the boson and fermion correlation functions at finite $T$, which will be used in the next section to calculate the charge transport at temperature $T$.
%In particular, we shall find that the disorder correlation length $\xi$ tunes the quasi-particle decay rate.
Crucially, the boson propagator develops a thermal gap at non-vanishing temperatures. The main contribution to this gap is from the four-boson interactions as previously discussed in \cite{Podolsky}. This contribution scales as $\Delta_b \approx \Pi_b(0) \sim \frac{U M_b T}{\log \Lambda/ T}$ \footnote{Although the fermions also contribute to the thermal boson gap, this contribution scales as $T^2/\Lambda_f$ and may be ignored}.
Therefore, the boson propagator at temperature $T$ has the form,
\begin{equation}
\begin{split}
    F(k,\omega_n) = \frac{1}{\Delta_b + \frac{k^2}{2M_b} + \alpha |\omega_n| - \im\eta\omega_n}\,.
    \label{eq:FT}
\end{split}
\end{equation}
where $\omega_n = 2\pi nT$ are the Matsubara frequencies. Analytical continuation of \eqref{eq:FT} to real frequencies results in the following retarded boson correlation function:
\begin{equation}
    F_R(k,\omega) = \frac{1}{\Delta_b + \frac{k^2}{2M_b} - \im\alpha\omega - \eta\omega}
    \label{eq:FR} \,.
\end{equation}

%The thermally gapped bosons do not instigate a full marginal Fermi liquid behavior anymore, and instead brings the fermions closer toward a Fermi-liquid. We demonstrate this by calculating the quasi-particle weight $Z$ and showing that it is close to $1$.
We now turn to the fermions and compute their retarded self-energy $\Sigma_R$;
from $\Sigma_R$ we can extract the quasi-particle relaxation rate $1/\tau$.
%The real-part of the self-energy $\Re\{\Sigma_R\}$ gives the fermion quasi-particle weight $Z$; together with the imaginary part of the self-energy $\Im\{\Sigma_R\}$, we find the relaxation rate through the relation, $\tau_k^{-1} = Z\Im\{\Sigma_R\}$

We first determine $\Re\{\Sigma_R\}$ and the corresponding fermion quasi-particle weight $Z$.
Due to the thermal boson gap, the Euclidean fermion self-energy deviates slightly from the marginal Fermi liquid behavior of \eqref{eq:SigmaE}.
At low frequencies $|\omega| < \frac{\Delta_b}\alpha$, it is linear in $\omega$, scaling as $\Sigma(k,\omega) \sim \im \gamma_k \log \left(\Lambda/\Delta_b\right) \omega$ (See Appendix.\ref{app:SD}).
Analytically continuing to real time, we find that the real part of $\Sigma_R$ is given as $\Sigma_R(k,\omega) \sim \gamma_k \log (\Lambda/\Delta_b) \omega \,.$
The corresponding quasi-particle weight $Z$ is,
\begin{equation}
Z = \frac{1}{1+\gamma_k \log \frac{\Lambda}{\Delta_b}} \,.
\label{eq:Z}
\end{equation}

$Z$ renormalizes the quasi-particle effective mass to $m_f/Z$. However, recall that the maximum value of $\gamma_k$ at most scales as $(k_F\xi)^{-1}$. This means that for moderate values of $k_F\xi$, the quasi-particle weight is essentially 1 at physically relevant temperatures. Thus, we may safely take the fermion mass to be its bare mass $m_f$.

%Since the quasi-particle weight $Z \simeq 1$, the quasi-particle decay rate $\tau_k$ is simply given by the imaginary part of the self-energy, $-\Im\{\Sigma_R(k,0)\}$.
We now turn to $\Im\{\Sigma_R(k,\omega)\}$.
By carrying out the convolution of the boson and fermion correlation functions in the Keldysh-framework (See Appendix.\ref{app:SD}), we obtain
\begin{equation}
\begin{split}
    & \Im \{\Sigma_R(k,\omega)\} \simeq  - \gamma_k T \Big(\lambda + \pi \log \cosh \frac{\beta\omega}{2}\Big) \,.
    %& f(\Omega,\Delta_b) = \int_\omega \left( \tan^{-1} \frac{\alpha \omega}{\Delta_b} - \tan^{-1}\frac{\alpha (\omega+\Omega)}{\Delta_b} \right) \tanh\frac{\beta\omega}{2} \,.
    \end{split}
    \label{eq:SigmaRimag}
\end{equation}
Here, $\lambda \sim \log \frac{\alpha T}{\Delta_b} = \log \log \frac{\Lambda}{T}$. Henceforth we shall treat it as an $O(1)$ constant.
%Here, $f$ is a dimensionless function of the dimensionless variables $\beta\Omega, \beta\Delta_b$. At $\Omega = 0$, $f(0,\beta\Delta_b) = 0$, and for large $\Omega \gg \frac{\Delta_b}{\alpha}$ asymptotically approaches $\pi \beta\Omega/2$.

Putting \eqref{eq:Z} and \eqref{eq:SigmaRimag} together, we find that $\tau_k^{-1}$, the scattering rate for a quasi-particle at momentum $k$ has the form,
\begin{equation}
    \tau_k^{-1} = Z \Im\{\Sigma_R\} \simeq \gamma_k T \,.
    \label{eq:decayrate}
\end{equation}
The scattering rate is most suppressed at the nodes, where it scales as $\tau_{\textrm{node}}^{-1} \sim (k_F\xi)^{-3} T$; elsewhere, it is much faster, scaling as $\tau_{k}^{-1} \sim \frac{\cos^2 2\theta_k}{k_F \xi} T$.

It is important to note that the scattering rate of \eqref{eq:decayrate} is different from the momentum relaxation rate.
After each collision, the momentum distribution of an excitation broadens by order $1/\xi$.
For complete momentum relaxation, the distribution has to spread over the entire Fermi surface.
This means that an excitation needs to undergo a number of collisions of the order of $(k_F \xi)^2$ to relax its momentum.
During this process, the nodes become a significant bottleneck due to the suppressed scattering rate in their vicinity. In the next section we use the quantum Boltzmann equation to account for the angular dependence of the scattering.
As we will show, the resulting overall momentum relaxation rate for an excitation scales as $1/\tau_{re} \sim (k_F \xi)^{-4} T$ and is notably sub-Planckian for $\xi \gg 1$.

\begin{comment}
Notably, the quasi-particle decay rate is linear in $T$, with a momentum dependent prefactor.
When $k_F \xi \gg 1$, $\gamma_k \ll 1$, and the decay rate is simply given as $\tau_k^{-1} \simeq \gamma_k T \ll T$ and overall is sub-Planckian.
In particular, at angles $(k_F \xi)^{-1}$ around the nodes, the decay rate is smallest, scaling as $\tau_{node}^{-1} \sim (k_F\xi)^{-3} T$; elsewhere, the decay is much faster, scaling as $\tau_{k}^{-1} \sim \frac{\cos^2 2\theta_k}{k_F \xi} T$.
As the disorder correlation length gets shorter, $\gamma_k$ gets larger and the decay rate gradually becomes faster, until it ultimately saturates at Planckian dissipation of $\tau_k^{-1} \simeq T$ when $k_F\xi \sim \BigO(1)$.
\end{comment}

\section{Transport}
\label{sec:dtransport}
In this section, we employ the quantum Boltzmann equation %together with the correlation functions and self-energies from the previous section 
to compute the transport properties.
In the DC limit it is given by \cite{mah00}
%Both the fermions and bosons are charged and contribute to charge transport in our model.
%As we will show, our model exhibits linear in $T$ resistivity at the critical point predominantly due to charge transport from unpaired fermions.
\begin{subequations}
\label{eq:BE0}
\begin{align}
    & A_F(q,\Omega)^2 \Im\{\Pi_R(q,\Omega)\} \frac{2e q \cdot E}{M_b} {b_0}'(\Omega) = \Pi^> F^< - \Pi^< F^> \,,
    \label{eq:BEb0} \\
    & A_G(k,\omega)^2 \Im\{\Sigma_R(k,\omega)\} \frac{e k \cdot E}{m_f} {f_0}'(\omega) = \Sigma^> G^< - \Sigma^< G^> \,.
    \label{eq:BEf0}
\end{align}
\end{subequations}
Here $f_0, b_0 = \frac{1}{e^{\beta\omega} \pm 1}$ are the Fermi-Dirac / Bose-Einstein distribution functions, and $A_{G,F}$ are the fermion/boson spectral functions. This 
equation takes the Green's functions computed in the previous section as input. It then describes the evolution of the generalized distribution functions in both momentum and frequency space in response to the external field, thereby allowing for calculation of transport even in the absence of sharp quasiparticles \cite{KimLeeWen,NaveLee,2004PhRvB..69c5111S,Maslov}. 
%Therefore, through the quantum Boltzmann equation we can accurately compute our model's charge conductivity despite the absence of quasi-particles.
%Charge transport is predominantly due to unpaired fermions,.
The equilibrium distributions of the fermions and bosons are given respectively by
$$n_{f}^0(k,\omega) = A_G(k,\omega) f_0(\omega), \ n_{b}^0(q,\Omega) = A_{F}(q,\Omega) b_0(\Omega)\,.$$

Inserting the Green's functions computed in the previous section we obtain the quantum Boltzmann equation for the deviations $\delta f(k,\omega), \delta b(q,\Omega)$ from the equilibrium distributions (See Appendix.\ref{app:QBE} for details),
\begin{widetext}
    \begin{subequations}
    \label{eq:BE}
    \begin{align}
        & A_F(q,\Omega)^2 \frac{2e q \cdot \vec E}{M_b} {b_0}'(\Omega) = \delta b(q,\Omega) \,,
        \label{eq:BEb} \\
        & \frac{e v_F \hat{k} \cdot \vec E f_0'(\omega)}{16\pi} = \int_{|\hat{k} + \hat{k'}| < \frac{\pi}{k_F\xi},\ \omega'}  \cos^2 2\theta_{\hat{k}-\hat{k'}} B(\omega+\omega') \big(1-f_{0}(\omega) \big) \big(1-f_0(\omega') \big) \{g_f(\hat{k},\omega) + g_f(\hat{k'},\omega')\} \,,
        \label{eq:BEf} \\
        &\textrm{ where } \delta f(\hat{k},\omega) = \int_{\Delta k} \delta f\big((k_F+\Delta k)\hat{k},\omega\big) = f_0(\omega) \big(1 - f_0(\omega) \big) g_f(\hat{k},\omega) \,, \ \ B(\omega) = b_{0}(\omega) \tan^{-1}\frac{\alpha \omega}{\Delta_b} \nonumber \,.
    \end{align}
    \end{subequations}
\end{widetext}
Here we have set $\hbar, k_B = 1$.
The left hand sides of \eqref{eq:BE} denote the evolution of the boson (fermion) distribution function in response to the external field. 
The right hand sides of the equations are the collision integrals that account for the relaxation processes due to the fermion-boson scattering processes.
Note that in the fermion quantum Boltzmann equation \eqref{eq:BEf}, we have used the fact that the fermion spectral function is still sharply peaked and integrated out the momentum direction perpendicular to the Fermi surface for simplicity.

The strategy to compute the conductivity from \eqref{eq:BE} is then the following: We first solve \eqref{eq:BE} for $\delta f$ and $\delta b$, from which we calculate the current of each species $J_{f,b}$.
The conductivity then naturally follows from the relation $\sigma = J/E$.
%; from these, we compute the current through the relation, $J_{f} = \int_{\hat{k},\omega} e k_F \hat{k} \delta f(\hat{k},\omega)$, $J_{b} = \int_{q,\Omega} \frac{2e q}{M_b} \delta b(q,\Omega)$. The conductivity naturally follows through the relation $\sigma = J/E$.

Let us start by computing $\sigma_b$, the boson contribution to the conductivity. The boson current is given by, $J_b = \int_{q,\Omega} \frac{2e q}{M_b} \delta b(q,\Omega)$. Dividing by the electric field, we find,
\begin{equation}
    \sigma_b^{xx} = \frac{e^2 \beta}{4M_b^2} \int_{q,\Omega} \left(q A_F(q,\Omega) \ \csch \frac{\beta\Omega}{2} \right)^2 \,.
    \label{eq:kubob_d} \\
\end{equation}
Evaluating the boson conductivity \eqref{eq:kubob_d} by inserting \eqref{eq:FT}, we find,
\begin{equation}
    \sigma_b^{xx} \simeq \frac{\alpha e^2}{M_b U}  \log \frac{\Lambda}{T} \,.
    \label{eq:bosoncond_d}
\end{equation}
Hence the Cooper pair fluctuations contribute a conductivity that scales logarithmically with temperature.

We now turn to the fermion conductivity, $\sigma_f$. We demonstrate through a scaling argument that it is inversely proportional to the temperature.
The key point is the following: the solution to \eqref{eq:BEf} for an electric field $\vec E = E \hat{x}$ can be written as $g(\hat{k},\omega) = e v_F \beta^2 E \ \tilde g(\hat{k},\tilde\omega = \beta\omega)$, where $\tilde g(\hat{k},\tilde\omega)$ is a dimensionless function independent of temperature (up to logarithmic corrections) that satisfies,
\begin{widetext}
\begin{equation}
\begin{split}
    \frac{\cos \theta_{\hat{k}} \tilde f_0(\tilde \omega) }{16\pi} &= \int_{\hat{k'},\tilde\omega'}^{|\hat{k} + \hat{k'}| < \frac\pi{k_F\xi}} \cos^2 2\theta_{\hat{k} - \hat{k'}}  \tilde B(\tilde\omega+\tilde\omega') \big(1 - \tilde f_0(\tilde\omega') \big) \big\{ \tilde g(\hat{k},\tilde\omega) + \tilde g(\hat{k'},\tilde\omega') \big\} \,, \\
    & \textrm{where } \tilde B(\tilde\omega) = \frac{\tan^{-1} \frac{\alpha \tilde\omega}{\tilde \Delta_b}}{e^{\tilde\omega}-1} , \ \ \tilde f_0(\tilde\omega) = \frac{1}{1+e^{\tilde\omega}} \,.
    \label{eq:BEf_rescaled}
\end{split}
\end{equation}
\end{widetext}
To see that $\tilde g$ is independent of temperature, notice that \eqref{eq:BEf_rescaled} is dependent only on dimensionless variables such as $\frac{1}{k_F\xi}$ and $\tilde \Delta_b = \frac{\Delta_b}{T} \sim \frac{U M_b}{\log \Lambda/T}$.
%To see that $\tilde g$ is independent of $T$, notice that in \eqref{eq:BEf_rescaled}, the function $\tilde f_0(\tilde \ome)$ is independent of temperature, and so is $\tilde B$.

Since $J_{f} = 2\int_{\hat{k},\omega} e k_F \hat{k} \delta f(\hat{k},\omega)$, we find that the longitudinal and Hall fermion conductivity are each given by,
\begin{equation}
\begin{split}
    \sigma_f^{xx} &=  \frac{2e^2\epsilon_F}{T} \int_{\hat{k},\tilde\omega} \cos \theta_k \tilde f_0(\tilde\omega) \big( 1 - \tilde f_0(\tilde\omega) \big) \tilde g(\hat{k},\tilde\omega) \,, \\
    \sigma_f^{xy} &=  \frac{2e^2\epsilon_F}{T} \int_{\hat{k},\tilde\omega} \sin \theta_k \tilde f_0(\tilde\omega) \big( 1 - \tilde f_0(\tilde\omega) \big) \tilde g(\hat{k},\tilde\omega) \,.
    \label{eq:sigmaf_d}
\end{split}
\end{equation}
Note that the integrals of the right hand sides of \eqref{eq:sigmaf_d} is independent of temperature:
It is therefore clear that the fermion conductivities are inversely proportional to $T$.

\begin{figure}[h]
    \centering
    \includegraphics[width = 0.9\columnwidth]{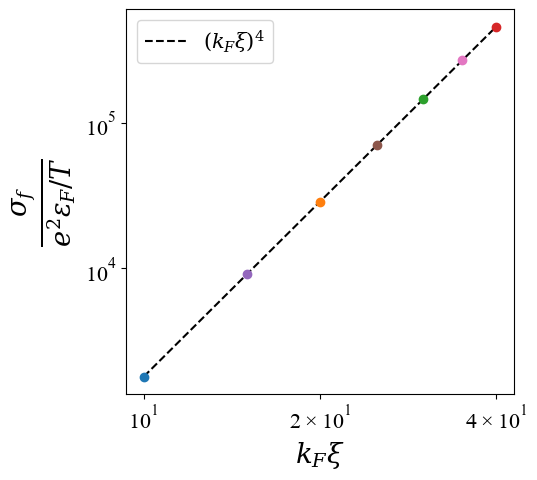}
    \caption{$\sigma_f T/e^2\epsilon_F$ vs $k_F\xi$ plotted in log-log scale for $\tilde \Delta_b = 1$, obtained from a numerical solution of the quantum Boltzmann equation, \eqref{eq:BEf}.}
    \label{fig:prefactor}
\end{figure}

To determine the proportionality factor in \eqref{eq:sigmaf_d}, we need to solve \eqref{eq:BEf_rescaled}.
This can be done numerically, and the result is displayed in Fig.~\ref{fig:prefactor}.
We find \footnote{$\sigma_f^{xy}$ is identically zero without a magnetic field present.},
\begin{equation}
    \sigma_f^{xx} \simeq 0.180 \ (k_F\xi)^4 \frac{e^2\epsilon_F}{T}
    \label{eq:fermioncond_d}
\end{equation}

This scaling behavior can be understood through a simple semiclassical argument.
We focus our attention on the bottleneck of the momentum relaxation process at the nodes.
In particular, let us take an excitation at an angle $\theta_k = \frac\pi4 + \phi$ without loss of generality.
Recall that the scattering rate at angle $\theta_k$ is $\tau_k$, and that after each collision the momentum changes at most by $\frac1{\xi}$.
This means that the rate at which its variance in momentum increases is given as,
$\partial_t {\braket{p^2}}_k \sim \xi^{-2}\tau_k^{-1}$.
As a result, the rate at which the angular variance increase is given by, $\partial_t \braket{\theta^2}_k \sim (k_F\xi)^{-2} \tau_k^{-1}$.
From this, we estimate the momentum relaxation time $\tau_{\textrm{tr}}$ as,
\begin{equation*}
    \tau_{\textrm{tr}} \simeq \int_{\theta_k} (k_F \xi)^2 \tau_k \simeq \int_{\phi} \frac{(k_F \xi)^5/T}{1+(k_F \xi)^2\phi^2} \sim (k_F\xi)^4/T\,.
    \label{eq:tau0}
\end{equation*}
Finally, using a Drude approximation of $\sigma_f \simeq \frac{ne^2 \tau_{\textrm{tr}}}{m}$, we arrive at,
\begin{equation*}
    \sigma_f \sim (k_F\xi)^4 \frac{e^2\epsilon_F}{T}
\end{equation*}
in agreement with the numerical solution of the quantum Boltzmann equations \eqref{eq:fermioncond_d}.
It follows that at low temperatures $\sigma_f \gg \sigma_b$. Hence the charge transport is dominated by the fermion contribution leading to $T$ linear resistivity

\begin{equation}
    \rho_{tot} \simeq \frac{5.56} {(k_F \xi)^{4}} \frac{T}{e^2 \epsilon_F}\,.
    \label{eq:rho}
\end{equation}

%The total conductivity $\sigma^{DC}$ scales as \eqref{eq:fermioncond_d} due to dominant charge transport from nodal quasi-particles. 

\subsection{Effects of Potential Disorder}
\label{subsec:ddd}

So far we have only considered the disorder in the boson-fermion pairing interaction. However it is natural to expect at least a small amount of regular single particle disorder coupling to the fermion density. Indeed, we argued in section  \ref{sec:model} that the long scale pairing disorder can be produced by short scale potential disorder. In this section we address the effect of the residual short scale potential disorder. Specifically we consider
%As we will explain shortly, the low temperature resistivity saturates to a nonzero constant in the presence of such disorder.
%However, the value that it saturates to is order $k_F\xi$ smaller than the value predicted from an extrapolation from the linear in $T$ regime.
%Furthermore, the constant low temperature and the linear in $T$ high temperature regimes are connected by an intermediate regime where the resistivity scales as $\sqrt{T}$.
%Both of these anomalous behaviors trace their origins to the strongly angle dependent quasi-particle decay rates.
the following random potential coupling to the fermion density:
\begin{equation}
    H_{rp} = V_{x} c^\dagger_x c_x\,, \textrm{ where } \overline{V_x} = 0, \ \overline{V_{x}V_{x'}} \sim W\delta_{x x'}\,.
    \label{eq:Hrp}
\end{equation}

Scattering off this random potential results in the following standard contribution to the fermion self-energy,
\begin{equation}
    \Sigma_{rp}(\omega) = \frac{\im\sgn(\omega)}{2\tau_0} \,, \textrm{ where } \frac1{\tau_0} = m_f W\,.
    \label{eq:Sigma_rp}
\end{equation}
Because the short range disorder scatters at all angles, unlike the interaction disorder, the above self energy is directly related to the momentum relaxation rate.

The self energy \eqref{eq:Sigma_rp} adds a contribution to the collision integral in the right hand side of the of the quantum Boltzmann equation  
\eqref{eq:BEf_rescaled} of $\frac{\tilde{f_0}(\tilde\omega)}{16\pi \tau_0 T} \tilde g(\hat{k},\tilde\omega)$ (See Appendix.\ref{app:QBE},\eqref{eq:appBEf_dd_rescaled}).
Consequently $\tilde g$, and therefore the conductance depend on $\tau_0 T$.

%Despite the additional self-energy of \eqref{eq:Sigma_rp} to the fermions, the boson dynamics are largely unaffected.
%As a consequence, the boson conductivity at the critical point is still given by \eqref{eq:bosoncond_d} and scales logarithmically with temperature.

%On the other hand, the fermion conductivity is greatly affected by the additional self-energy of \eqref{eq:Sigma_rp}.
%In the quantum Boltzmann equation, the effect of the random potential manifests itself in an extra term in the right hand side (the collision integral) of 
In particular, in the limit $T\to 0$ the potential disorder leads to a saturation of the resistivity to a non vanishing constant
\begin{equation}
    \rho_0 = \frac{2\pi}{e^2 \epsilon_F \tau_0} \,.
    \label{eq:rhoddd_lowT}
\end{equation}
Note that for weak potential disorder $\tau_0$ is large and $\rho_{0}$ can be several orders of magnitude smaller than the resistance quantum $R_Q$.

\begin{figure}
    \centering
    \includegraphics[width = 0.90\columnwidth]{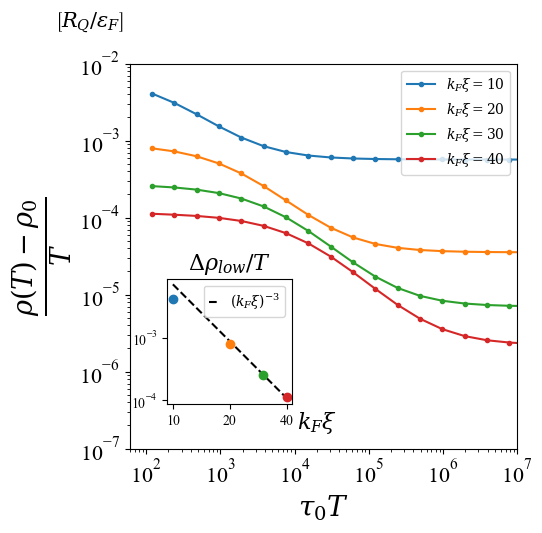}
    \caption{Log-log plot of $\frac{\rho(T)-\rho_0}T$ versus temperature in the presence of potential disorder for various disorder correlation lengths. The temperature axis has been rescaled by the electron mean free time $\tau_0$. At high temperatures $\tau_0 T \gg (k_F\xi)^4$, the resistivity scales linearly in $T$ as given by \eqref{eq:rho}.
    At low temperatures resistivity saturates to a constant $\rho_{0}$ \eqref{eq:rhoddd_lowT}.
    Additionally, resistivity increases linearly with $T$ at low temperatures, albeit with a different slope than at high $T$.
    The inset graph shows how this slope scales with the disorder correlation length. It scales as, $\frac{\Delta \rho_{low}}{T} \sim \frac{(k_F\xi)^{-3}}{\epsilon_F}$.}
    \label{fig:ddlow}
\end{figure}

Comparing the two contributions to the collision integral suggests a crossover at a temperature scale $T \gg T_0= \frac{(k_F\xi)^4}{\tau_0}$ between the
low temperature limit affected by the potential disorder to the higher temperature regime dominated by the interaction disorder. As expected, the resistivity for $T\gg T_0$ approaches the result \ref{eq:rho}. On the other hand the leading temperature dependence for $T\ll T_0$ can be obtained by 
solving the quantum Boltzmann equation perturbatively around the zero temperature solution. This gives
\begin{equation}
    \rho_{low} - \rho_0 \simeq 16.8 (k_F\xi)^{-3} \frac{T}{e^2 \epsilon_F}\,.
    \label{eq:rholow}
\end{equation}
That is, we find linear in $T$ resistivity also in the low temperature limit, but with a larger slope compared to the regime $T>T_0$.
The reason for the stronger temperature dependence is that the scattering from the potential disorder eliminates the nodal bottleneck for relaxation.
As a result, scattering off of Cooper pairs contribute a faster relaxation rate of $(k_F\xi)^{-2} \tau_{\textrm{antinode}}^{-1} \sim (k_F\xi)^{-3}T$ that is reflected in \eqref{eq:rholow}.
The resistivity obtained from numerical solution  
of the quantum Boltzmann equation, 
shown in Fig.\ref{fig:ddlow}, corroborates the analytic results.
%Solving the Boltzmann equation perturbatively around the zero temperature solution gives the leading temperature dependence in the low temperature limit. Interestingly this is found to be linear in $T$ as 

%At high temperatures $T \gg T_0= \frac{(k_F\xi)^4}{\tau_0}$, the scattering on Cooper pairs through the disordered interaction dominates and the resistivity tends back to the form  \eqref{eq:rho}. 

\subsection{Magnetotransport}
\label{subsec:MR}

We have seen above that the fermions dominate the transport. We consider their contribution to the magnetoresistance first. We then confirm that the boson contribution is negligible.  

The quantum Boltzmann equation for the fermion distribution is modified by a perpendicular magnetic field, bringing the left hand side of\eqref{eq:BEf_rescaled} 
into the following form
\begin{equation}
    \frac{\cos\theta_k \tilde f_0(\tilde\omega)}{16\pi} - \frac{\omega_{cf}/T}{16\pi} \tilde{f}_0(\tilde\omega) \partial_{\theta_k} g(\hat{k},\tilde\omega) \,,
    \label{eq:BEf_rescaled_B}
\end{equation}
where $\omega_{cf}=eB/m_f$ is the fermion cyclotron frequency. We see that the magnetic field and the temperature appear in this scaled equation only through the dimensionless ratio $b=\omega_{cf}/T$ and therefore the solution $\tilde g$ must also depend only on $b$. The conductivity, determined from  $\tilde g$ through the relation \eqref{eq:sigmaf_d} must then take the scaling form $\sigma =e^2\epsilon_F/T \,\phi(\omega_{cf}/T)$. Similarly for the magnetoresistance $\rho(T,B)-\rho(T,0)= (T/e^2\epsilon_F)\psi(\omega_{cf}/T)$. %, and ultimately the longitudinal resistance divided by the temperature, $\frac{\rho(B,T)}{T}$ also.

\begin{figure}
    \centering
    \includegraphics[width = 0.95\columnwidth]{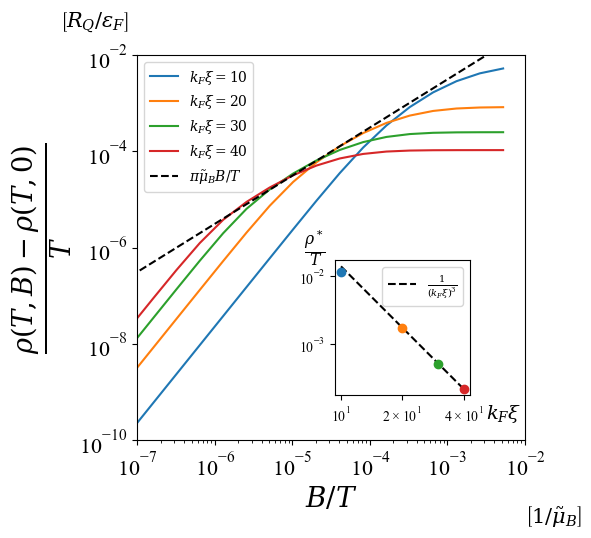}
    \caption{Log-log plot of $B/T$ vs $\frac{\rho(T,B)-\rho(T,0)}{T}$ for various disorder correlation lengths. For magnetic fields such that $(k_F \xi)^{-5} \ll b=\frac{\tilde\mu_B B}{T} \ll (k_F \xi)^{-3}$, longitudinal resistivity is linear in $B$ with slope $\frac{\pi \tilde \mu_B}{\epsilon_F}$ (Dashed line), where $\tilde\mu_B = \frac{e}{m_f}$ denote the renormalized bohr magneton. For weak fields of $b \ll (k_F\xi)^{-5}$, resistivity behaves as \eqref{eq:fermioncond_d}. Inset shows that for strong fields $b \gg (k_F \xi)^{-3}$, resistivity saturates to scaling as $\rho^* \sim (k_F \xi)^{-3} \frac{T}{e^2 \epsilon_F}$.}
    \label{fig:sigf_B}
\end{figure}

Fig.\ref{fig:sigf_B} shows the behavior of the magnetoresistance by plotting the scaling function $\psi(b)$ obtained from numerical solution of the quantum Boltzmann equation for different values of the scaled disorder correlation length $k_f\xi$. We find that the magnetoresistance is linear in $B$ within the range of magnetic fields such that $T(k_F\xi)^{-4} \ll \omega_{cf} \ll T(k_F\xi)^{-3}$. Note that the slope of the linear $B$ dependence is independent of the disorder (i.e. of $\xi$) and is given  by   $\pi \tilde\mu_B /(e^2 \epsilon_F)$, where $\tilde \mu_B = \frac{e}{m_f}$.

The linear in field magnetoresistance arises from the strongly anisotropic quasi-particle scattering rate \eqref{eq:decayrate} as we now explain.
Recall that in absence of a magnetic field the nodes at $\theta_k = \pm\frac{\pi}{4}, \pm\frac{3\pi}{4}$ act bottlenecks for momentum relaxation.
Once a magnetic field is turned on, quasiparticles begin to rotate around the Fermi surface at a rate $\omega_{cf}$. This  can overcome the nodal bottleneck if the rotation is faster than the rate at which momentum relaxes at the nodes, i.e. $\omega_{cf} \gg (k_F\xi)^{-4} T$. In this case the time scale for momentum relaxation rate can be replaced by $\omega_{cf} = \tilde\mu_B B$, leading to linear magneto resistance. However if $\omega_{cf}$ exceeds the rate of momentum relaxation away from the nodes, i.e. $\omega_{cf}\gg T(k_F\xi)^{-3}$, then this anti-nodal scattering becomes the bottleneck for momentum relaxation and the resistivity saturates to 
%In turn, the transport lifetime $\tau_{\textrm{tr}} \sim \frac1{\omega_{cf}}$ is linear in $B$ with a coefficient of $\tilde \mu_B = \frac{e}{m_f}$, and a Drude approximation results in a linear magnetoresistance of,
%\begin{equation*}
%    \rho(B) \simeq \frac{m_f \tilde \mu_B B}{ne^2} \sim \frac{\tilde \mu_B B}{e^2\epsilon_F} \,.
%\end{equation*}
$\rho^* \simeq (k_F\xi)^{-3} T/(e^2\epsilon_F)$.

We can now explain why the contribution of the bosons to the magnetoresistance is negligible compared with that of the fermions. In the regime with linear magnetoresistance the fermion conductivity is reduced from $e^2 (\epsilon_F/T) (k_F\xi)^4$ to $e^2 (\epsilon_F/T) (k_F\xi)^3$. This change is much larger than the entire boson contribution to the conductivity, which is, up to logarithmic corrections, just of order of the resistance quantum.
As a consequence, any changes in the boson conductivity is washed out by the change in fermion conductivity.
We leave a more detailed derivation of the boson magnetotransport to Appendix.\ref{app:QBE}.

\section{Diamagnetic Susceptibility}
\label{sec:DS}
%In this section, we make a striking prediction of the diamagnetic susceptibility $\chi$ that could be measured in experiments to verify the physics of our model.
%As we will show, $\chi$ diverges logarithmically with temperature at the critical point in our model.
%This divergence signifies the onset of the Magnus effect due to superconductivity beyond the critical point.
%However, the slow logarithmic divergence is in sharp contrast to that of standard XY transitions, where $\chi$ diverges as $1/T$.
%Such clear distinction allows for experimental measurements of the diamagnetic susceptibility to serve as a test of our model.

The diamagnetic susceptibility can be calculated from the transverse current-current correlator at $\omega=0$
\begin{equation}    \chi=\lim_{q\to0}\partial^2_q\left\langle J_x(q\hat{y},0)J_x(-q\hat{y},0)\right\rangle.
\end{equation}
The scaling of $\chi$ with temperature at the critical point provides a further test of our theory of the QSMT. 

The diamagnetic response is a sum of the fermionic and bosonic  contributions
given by
\begin{subequations}
\begin{align}
    \chi_f &= -\frac{2e^2 T}{(2\pi)^2} \partial_{q}^2 \sum_{\omega_n} \int_k \frac{k_x^2}{m_f^2} \left[ G_{k+\frac{q\hat{y}}2}(\omega_n) G_{k-\frac{q\hat{y}}2}(\omega_n)  \right] \,,
    \label{eq:chifog} \\
    \chi_b &= -\frac{4e^2 T}{(2\pi)^2}  \partial_{q}^2 \sum_{\omega_n} \int_k \frac{k_x^2}{M_b^2} \left[ F_{k+\frac{q\hat{y}}2}(\omega_n) F_{k-\frac{q\hat{y}}2}(\omega_n) \right] \,,
    \label{eq:chibog}
\end{align}
\end{subequations}
in the limit of $q \rightarrow 0$.

The fermions contribution gives essentially the standard, temperature independent Landau diamagnetic response (See Appendix.\ref{app:dia}),
\begin{equation}
    \chi_f \simeq \frac{e^2}{12\pi m_f}\,.
    \label{eq:chif}
\end{equation}
The bosons on the other hand give rise to a singular contribution at the critical point.  Using the boson propagator \eqref{eq:FT} in 
\eqref{eq:chibog} and integrating over the momentum gives
\begin{equation*}
    \chi_b = \frac{e^2}{6\pi \beta}\sum_{\omega_n} \frac{1}{\Delta_b + \alpha|\omega_n|} \,.
\end{equation*}
The contribution to this sum fron the zeroth Matsubara frequency diverges logarithmically with temperature because $\Delta_b \sim U M_b T/\log \frac{\Lambda}{T}$). The sum over the non-zero frequencies also gives a logarithmic divergence. So, together we obtain
\begin{equation}
    \chi_b \simeq \left(\frac{e^2}{U M_b} + \frac{e^2}{12\pi^2 \alpha} \right) \log\frac{\Lambda}{T}
    \label{eq:chib}
\end{equation}
We conclude that the total diamagnetic response diverges logarithmically with temperature at the critical point.
Note that whereas the charge transport was dominated by the fermionic quasiparticles, the critical diamagnetic response is   governed by the critical bosonic (Cooper pair) fluctuations.

The logarithmically divergent  susceptibility in our model should be compared with the  diamagnetic response in a standard XY transition \cite{doniach}. Within a Gaussian theory the result is $\chi\sim 1/T$ for the quantum critical point in two spatial dimensions. This would be modified to a non-trivial power-law in the true (Willson-Fisher) fixed point in 2+1 dimensdions, markedly different from the logarithmic temperature dependece we obtain. 

\section{Summary and Discussion}\label{sec:discussion}

We presented a theory for an unusual quantum superconductor to metal transition (QSMT) observed in nano-patterned YBCO films \cite{2022Natur.601..205Y}.
The film in this experiment exhibits $T$-linear resistivity with very small residual resistivity at the QSMT tuned by the size of nano-patterned holes. This stands in contrast to the Fermi liquid phenomenology characterizing the superconductor to metal transition obtained in cuprates by doping Zinc impurities \cite{Fukuzumi} or at the overdoped edge of the superconducting dome \cite{2009Sci...323..603C}.
The essential ingredient in our model, leading to the destruction of quasi-particles and the $T$-linear resistivity, is a disordered and frustrated coupling between the quasipaticles and the critical bosonic fluctuations (i.e. $d$-wave Cooper pairs).
Results of mean field calculations \cite{dunghai} provide indirect evidence that such effective interactions can emerge at intermediate scales due to the interplay of d-wave pairing and a disorder potential.
Our results show that $T$ linear resistivity can be obtained over a broad range if the emergent disordered interaction is stronger than the usual disordered potential coupling to the fermion density.
It is an open question how this situation emerges from the microscopics, but it is possible that patterning the cuprate film with holes, as done in Ref. \cite{2022Natur.601..205Y}, arranges the required conditions. A direct microscopic derivation of the effective interactions remain a challenge for future work.

%Our model also displays at the critical point a linear magnetoresistance (LMR).
%This phenomenon arose from a nodal bottleneck in momentum relaxation being replaced with the cyclotron frequency $\frac{eB}{m_f}$ in the presence of a magnetic field.
%Nevertheless, LMR arises at too weak a field strength within our model to explain the experiments.
%In a follow-up work, we demonstrate that more robust linear magnetoresistance can occur in the vicinity of order, where the order parameter either has a finite wavevector or has nodes \cite{shubhayu}.

Within the effective model we studied, the dominant contribution to the conductivity, scaling as $1/T$, comes from fermionic quasi-particles decaying into Landau damped critical Cooper pairs. The critical bosons themselves contribute only a sub-leading logarithmic divergence $\sigma_b \sim \log \Lambda/T$. In contrast, the authors of the experimental paper \cite{2022Natur.601..205Y} argue that the transport is dominated by the critical bosons. To support this interpretation they show oscillations of the magnetoresistance with a period set by the superconducting flux quantum $h/2e$ through a unit cell of the patterned superlattice. We note, however, that such oscillations can also arise from the the sub-leading contribution of the bosons in our model. As we show in Appendix \ref{app:LittleParks}, the temperature dependence of the magnetoresistance oscillation shown in Ref. \cite{2022Natur.601..205Y}, is consistent with the $\log (\Lambda/T)$ behavior of the bosonic contribution to the resistivity in our model. 

It is worth noting that similar phenomenology has been observed  in nano-patterned films of the iron based superconductor FeSe \cite{YananLiPreprint}. While here we focused on $d$-wave pairing, our model can be easily adapted to describe a QSMT involving other pairing symmetries. In particular, interplay of a random potential and the sign structure of the $s_\pm$ pairing wave-function proposed for this superconductor can lead to a disordered sign-changing pairing interaction through the same mechanism as in $d$-wave superconductors.
The transport calculations would be almost identical except for minor differences due to the absence of nodes in the coupling of Fermi surface quasi-particles to the critical fluctuations. Because the nodal bottleneck for momentum relaxation is removed, we expect $\rho\sim 
(k_F \xi)^{-3} \epsilon_F/T$ in the nodeless case versus $\rho\sim (k_F \xi)^{-4} \epsilon_F/T$ for a nodal pairing interaction. 

Finally, having discussed the critical point with T linear resistivity, it is worth noting that the samples which we regarded as being on the superconducting side of it in Ref. \cite{2022Natur.601..205Y} are not truly superconducting. At the lowest measured temperatures they behave as failed superconductors (or anomalous metals) with very low resistivity \cite{AMrmp}). Developing an effective description of this phase is an interesting problem for future work.

\begin{acknowledgments}
We would like to thank Yimu Bao, Xiangyu Cao, Shubhayu Chatterjee, Debanjan Chowdhury, Hart Goldman, Tobias Holder, Ruihua Fan, Chunxiao Liu, Aavishkar Patel, Xuepeng Wang, and Zack Weinstein for helpful discussions. E.B. acknowledges support from NSF-BSF award DMR-2000987 and from the European Research Council (ERC) under grant HQMAT (grant agreement No. 817799). Part of this work was done in the Aspen Center for Physics, supported by National Science Foundation grant PHY-2210452.\end{acknowledgments}

\bibliography{ref}

\begin{thebibliography}{38}%
\makeatletter
\providecommand \@ifxundefined [1]{%
 \@ifx{#1\undefined}
}%
\providecommand \@ifnum [1]{%
 \ifnum #1\expandafter \@firstoftwo
 \else \expandafter \@secondoftwo
 \fi
}%
\providecommand \@ifx [1]{%
 \ifx #1\expandafter \@firstoftwo
 \else \expandafter \@secondoftwo
 \fi
}%
\providecommand \natexlab [1]{#1}%
\providecommand \enquote  [1]{``#1''}%
\providecommand \bibnamefont  [1]{#1}%
\providecommand \bibfnamefont [1]{#1}%
\providecommand \citenamefont [1]{#1}%
\providecommand \href@noop [0]{\@secondoftwo}%
\providecommand \href [0]{\begingroup \@sanitize@url \@href}%
\providecommand \@href[1]{\@@startlink{#1}\@@href}%
\providecommand \@@href[1]{\endgroup#1\@@endlink}%
\providecommand \@sanitize@url [0]{\catcode `\\12\catcode `\$12\catcode
  `\&12\catcode `\#12\catcode `\^12\catcode `\_12\catcode `\%12\relax}%
\providecommand \@@startlink[1]{}%
\providecommand \@@endlink[0]{}%
\providecommand \url  [0]{\begingroup\@sanitize@url \@url }%
\providecommand \@url [1]{\endgroup\@href {#1}{\urlprefix }}%
\providecommand \urlprefix  [0]{URL }%
\providecommand \Eprint [0]{\href }%
\providecommand \doibase [0]{http://dx.doi.org/}%
\providecommand \selectlanguage [0]{\@gobble}%
\providecommand \bibinfo  [0]{\@secondoftwo}%
\providecommand \bibfield  [0]{\@secondoftwo}%
\providecommand \translation [1]{[#1]}%
\providecommand \BibitemOpen [0]{}%
\providecommand \bibitemStop [0]{}%
\providecommand \bibitemNoStop [0]{.\EOS\space}%
\providecommand \EOS [0]{\spacefactor3000\relax}%
\providecommand \BibitemShut  [1]{\csname bibitem#1\endcsname}%
\let\auto@bib@innerbib\@empty
%</preamble>
\bibitem [{\citenamefont {{Warren}}\ \emph {et~al.}(1989)\citenamefont
  {{Warren}}, \citenamefont {{Walstedt}}, \citenamefont {{Brennert}},
  \citenamefont {{Cava}},\ and\ \citenamefont {{Tycko}}}]{Pseudogap1}%
  \BibitemOpen
  \bibfield  {author} {\bibinfo {author} {\bibfnamefont {Jr.}\ \bibnamefont
  {{Warren}}, \bibfnamefont {W.~W.}}, \bibinfo {author} {\bibfnamefont {R.~E.}\
  \bibnamefont {{Walstedt}}}, \bibinfo {author} {\bibfnamefont {G.~F.}\
  \bibnamefont {{Brennert}}}, \bibinfo {author} {\bibfnamefont {R.~J.}\
  \bibnamefont {{Cava}}}, \ and\ \bibinfo {author} {\bibfnamefont
  {R.}~\bibnamefont {{Tycko}}},\ }\bibfield  {title} {\enquote {\bibinfo
  {title} {{Cu spin dynamics and superconducting precursor effects in planes
  above T$_{c}$ in YBa$_{2}$Cu$_{3}$O$_{6.7}$}},}\ }\href {\doibase
  10.1103/PhysRevLett.62.1193} {\bibfield  {journal} {\bibinfo  {journal}
  {\prl}\ }\textbf {\bibinfo {volume} {62}},\ \bibinfo {pages} {1193--1196}
  (\bibinfo {year} {1989})}\BibitemShut {NoStop}%
\bibitem [{\citenamefont {{Alloul}}\ \emph {et~al.}(1989)\citenamefont
  {{Alloul}}, \citenamefont {{Ohno}},\ and\ \citenamefont
  {{Mendels}}}]{Pseudogap2}%
  \BibitemOpen
  \bibfield  {author} {\bibinfo {author} {\bibfnamefont {H.}~\bibnamefont
  {{Alloul}}}, \bibinfo {author} {\bibfnamefont {T.}~\bibnamefont {{Ohno}}}, \
  and\ \bibinfo {author} {\bibfnamefont {P.}~\bibnamefont {{Mendels}}},\
  }\bibfield  {title} {\enquote {\bibinfo {title} {{$^{89}$Y NMR evidence for a
  fermi-liquid behavior in YBa$_{2}$Cu$_{3}$O$_{6+x}$}},}\ }\href {\doibase
  10.1103/PhysRevLett.63.1700} {\bibfield  {journal} {\bibinfo  {journal}
  {\prl}\ }\textbf {\bibinfo {volume} {63}},\ \bibinfo {pages} {1700--1703}
  (\bibinfo {year} {1989})}\BibitemShut {NoStop}%
\bibitem [{\citenamefont {{Doiron-Leyraud}}\ \emph {et~al.}(2007)\citenamefont
  {{Doiron-Leyraud}}, \citenamefont {{Proust}}, \citenamefont {{Leboeuf}},
  \citenamefont {{Levallois}}, \citenamefont {{Bonnemaison}}, \citenamefont
  {{Liang}}, \citenamefont {{Bonn}}, \citenamefont {{Hardy}},\ and\
  \citenamefont {{Taillefer}}}]{Pseudogap3}%
  \BibitemOpen
  \bibfield  {author} {\bibinfo {author} {\bibfnamefont {Nicolas}\ \bibnamefont
  {{Doiron-Leyraud}}}, \bibinfo {author} {\bibfnamefont {Cyril}\ \bibnamefont
  {{Proust}}}, \bibinfo {author} {\bibfnamefont {David}\ \bibnamefont
  {{Leboeuf}}}, \bibinfo {author} {\bibfnamefont {Julien}\ \bibnamefont
  {{Levallois}}}, \bibinfo {author} {\bibfnamefont {Jean-Baptiste}\
  \bibnamefont {{Bonnemaison}}}, \bibinfo {author} {\bibfnamefont {Ruixing}\
  \bibnamefont {{Liang}}}, \bibinfo {author} {\bibfnamefont {D.~A.}\
  \bibnamefont {{Bonn}}}, \bibinfo {author} {\bibfnamefont {W.~N.}\
  \bibnamefont {{Hardy}}}, \ and\ \bibinfo {author} {\bibfnamefont {Louis}\
  \bibnamefont {{Taillefer}}},\ }\bibfield  {title} {\enquote {\bibinfo {title}
  {{Quantum oscillations and the Fermi surface in an underdoped high-T$_{c}$
  superconductor}},}\ }\href {\doibase 10.1038/nature05872} {\bibfield
  {journal} {\bibinfo  {journal} {\nat}\ }\textbf {\bibinfo {volume} {447}},\
  \bibinfo {pages} {565--568} (\bibinfo {year} {2007})},\ \Eprint
  {http://arxiv.org/abs/0801.1281} {arXiv:0801.1281 [cond-mat.supr-con]}
  \BibitemShut {NoStop}%
\bibitem [{\citenamefont {Gurvitch}\ and\ \citenamefont
  {Fiory}(1987)}]{Cuprate1}%
  \BibitemOpen
  \bibfield  {author} {\bibinfo {author} {\bibfnamefont {M.}~\bibnamefont
  {Gurvitch}}\ and\ \bibinfo {author} {\bibfnamefont {A.~T.}\ \bibnamefont
  {Fiory}},\ }\bibfield  {title} {\enquote {\bibinfo {title} {Resistivity of
  ${\mathrm{la}}_{1.825}$${\mathrm{sr}}_{0.175}$${\mathrm{cuo}}_{4}$ and
  ${\mathrm{yba}}_{2}$${\mathrm{cu}}_{3}$${\mathrm{o}}_{7}$ to 1100 k: Absence
  of saturation and its implications},}\ }\href {\doibase
  10.1103/PhysRevLett.59.1337} {\bibfield  {journal} {\bibinfo  {journal}
  {Phys. Rev. Lett.}\ }\textbf {\bibinfo {volume} {59}},\ \bibinfo {pages}
  {1337--1340} (\bibinfo {year} {1987})}\BibitemShut {NoStop}%
\bibitem [{\citenamefont {Takagi}\ \emph {et~al.}(1992)\citenamefont {Takagi},
  \citenamefont {Batlogg}, \citenamefont {Kao}, \citenamefont {Kwo},
  \citenamefont {Cava}, \citenamefont {Krajewski},\ and\ \citenamefont
  {Peck}}]{Cuprate2}%
  \BibitemOpen
  \bibfield  {author} {\bibinfo {author} {\bibfnamefont {H.}~\bibnamefont
  {Takagi}}, \bibinfo {author} {\bibfnamefont {B.}~\bibnamefont {Batlogg}},
  \bibinfo {author} {\bibfnamefont {H.~L.}\ \bibnamefont {Kao}}, \bibinfo
  {author} {\bibfnamefont {J.}~\bibnamefont {Kwo}}, \bibinfo {author}
  {\bibfnamefont {R.~J.}\ \bibnamefont {Cava}}, \bibinfo {author}
  {\bibfnamefont {J.~J.}\ \bibnamefont {Krajewski}}, \ and\ \bibinfo {author}
  {\bibfnamefont {W.~F.}\ \bibnamefont {Peck}},\ }\bibfield  {title} {\enquote
  {\bibinfo {title} {Systematic evolution of temperature-dependent resistivity
  in
  ${\mathrm{la}}_{2\mathrm{\ensuremath{-}}\mathit{x}}$${\mathrm{sr}}_{\mathit{x}}$${\mathrm{cuo}}_{4}$},}\
  }\href {\doibase 10.1103/PhysRevLett.69.2975} {\bibfield  {journal} {\bibinfo
   {journal} {Phys. Rev. Lett.}\ }\textbf {\bibinfo {volume} {69}},\ \bibinfo
  {pages} {2975--2978} (\bibinfo {year} {1992})}\BibitemShut {NoStop}%
\bibitem [{\citenamefont {{Taillefer}}(2010)}]{Cuprate3}%
  \BibitemOpen
  \bibfield  {author} {\bibinfo {author} {\bibfnamefont {Louis}\ \bibnamefont
  {{Taillefer}}},\ }\bibfield  {title} {\enquote {\bibinfo {title} {{Scattering
  and Pairing in Cuprate Superconductors}},}\ }\href {\doibase
  10.1146/annurev-conmatphys-070909-104117} {\bibfield  {journal} {\bibinfo
  {journal} {Annual Review of Condensed Matter Physics}\ }\textbf {\bibinfo
  {volume} {1}},\ \bibinfo {pages} {51--70} (\bibinfo {year} {2010})},\ \Eprint
  {http://arxiv.org/abs/1003.2972} {arXiv:1003.2972 [cond-mat.supr-con]}
  \BibitemShut {NoStop}%
\bibitem [{\citenamefont {Fukuzumi}\ \emph {et~al.}(1996)\citenamefont
  {Fukuzumi}, \citenamefont {Mizuhashi}, \citenamefont {Takenaka},\ and\
  \citenamefont {Uchida}}]{Fukuzumi}%
  \BibitemOpen
  \bibfield  {author} {\bibinfo {author} {\bibfnamefont {Y.}~\bibnamefont
  {Fukuzumi}}, \bibinfo {author} {\bibfnamefont {K.}~\bibnamefont {Mizuhashi}},
  \bibinfo {author} {\bibfnamefont {K.}~\bibnamefont {Takenaka}}, \ and\
  \bibinfo {author} {\bibfnamefont {S.}~\bibnamefont {Uchida}},\ }\bibfield
  {title} {\enquote {\bibinfo {title} {Universal superconductor-insulator
  transition and ${T}_{c}$ depression in zn-substituted high- ${T}_{c}$
  cuprates in the underdoped regime},}\ }\href {\doibase
  10.1103/PhysRevLett.76.684} {\bibfield  {journal} {\bibinfo  {journal} {Phys.
  Rev. Lett.}\ }\textbf {\bibinfo {volume} {76}},\ \bibinfo {pages} {684--687}
  (\bibinfo {year} {1996})}\BibitemShut {NoStop}%
\bibitem [{\citenamefont {Hebard}\ and\ \citenamefont
  {Paalanen}(1990)}]{Hebard1}%
  \BibitemOpen
  \bibfield  {author} {\bibinfo {author} {\bibfnamefont {A.~F.}\ \bibnamefont
  {Hebard}}\ and\ \bibinfo {author} {\bibfnamefont {M.~A.}\ \bibnamefont
  {Paalanen}},\ }\bibfield  {title} {\enquote {\bibinfo {title}
  {Magnetic-field-tuned superconductor-insulator transition in two-dimensional
  films},}\ }\href {\doibase 10.1103/PhysRevLett.65.927} {\bibfield  {journal}
  {\bibinfo  {journal} {Phys. Rev. Lett.}\ }\textbf {\bibinfo {volume} {65}},\
  \bibinfo {pages} {927--930} (\bibinfo {year} {1990})}\BibitemShut {NoStop}%
\bibitem [{\citenamefont {Paalanen}\ \emph {et~al.}(1992)\citenamefont
  {Paalanen}, \citenamefont {Hebard},\ and\ \citenamefont {Ruel}}]{Hebard2}%
  \BibitemOpen
  \bibfield  {author} {\bibinfo {author} {\bibfnamefont {M.~A.}\ \bibnamefont
  {Paalanen}}, \bibinfo {author} {\bibfnamefont {A.~F.}\ \bibnamefont
  {Hebard}}, \ and\ \bibinfo {author} {\bibfnamefont {R.~R.}\ \bibnamefont
  {Ruel}},\ }\bibfield  {title} {\enquote {\bibinfo {title} {Low-temperature
  insulating phases of uniformly disordered two-dimensional superconductors},}\
  }\href {\doibase 10.1103/PhysRevLett.69.1604} {\bibfield  {journal} {\bibinfo
   {journal} {Phys. Rev. Lett.}\ }\textbf {\bibinfo {volume} {69}},\ \bibinfo
  {pages} {1604--1607} (\bibinfo {year} {1992})}\BibitemShut {NoStop}%
\bibitem [{\citenamefont {Seidler}\ \emph {et~al.}(1992)\citenamefont
  {Seidler}, \citenamefont {Rosenbaum},\ and\ \citenamefont {Veal}}]{Seidler}%
  \BibitemOpen
  \bibfield  {author} {\bibinfo {author} {\bibfnamefont {G.~T.}\ \bibnamefont
  {Seidler}}, \bibinfo {author} {\bibfnamefont {T.~F.}\ \bibnamefont
  {Rosenbaum}}, \ and\ \bibinfo {author} {\bibfnamefont {B.~W.}\ \bibnamefont
  {Veal}},\ }\bibfield  {title} {\enquote {\bibinfo {title} {Two-dimensional
  superconductor-insulator transition in bulk single-crystal
  ${\mathrm{yba}}_{2}$${\mathrm{cu}}_{3}$${\mathrm{o}}_{6.38}$},}\ }\href
  {\doibase 10.1103/PhysRevB.45.10162} {\bibfield  {journal} {\bibinfo
  {journal} {Phys. Rev. B}\ }\textbf {\bibinfo {volume} {45}},\ \bibinfo
  {pages} {10162--10164} (\bibinfo {year} {1992})}\BibitemShut {NoStop}%
\bibitem [{\citenamefont {van~der Zant}\ \emph {et~al.}(1992)\citenamefont
  {van~der Zant}, \citenamefont {Fritschy}, \citenamefont {Elion},
  \citenamefont {Geerligs},\ and\ \citenamefont {Mooij}}]{Zant}%
  \BibitemOpen
  \bibfield  {author} {\bibinfo {author} {\bibfnamefont {H.~S.~J.}\
  \bibnamefont {van~der Zant}}, \bibinfo {author} {\bibfnamefont {F.~C.}\
  \bibnamefont {Fritschy}}, \bibinfo {author} {\bibfnamefont {W.~J.}\
  \bibnamefont {Elion}}, \bibinfo {author} {\bibfnamefont {L.~J.}\ \bibnamefont
  {Geerligs}}, \ and\ \bibinfo {author} {\bibfnamefont {J.~E.}\ \bibnamefont
  {Mooij}},\ }\bibfield  {title} {\enquote {\bibinfo {title} {Field-induced
  superconductor-to-insulator transitions in josephson-junction arrays},}\
  }\href {\doibase 10.1103/PhysRevLett.69.2971} {\bibfield  {journal} {\bibinfo
   {journal} {Phys. Rev. Lett.}\ }\textbf {\bibinfo {volume} {69}},\ \bibinfo
  {pages} {2971--2974} (\bibinfo {year} {1992})}\BibitemShut {NoStop}%
\bibitem [{\citenamefont {Wang}\ \emph {et~al.}(1991)\citenamefont {Wang},
  \citenamefont {Beauchamp}, \citenamefont {Berkley}, \citenamefont {Johnson},
  \citenamefont {Liu}, \citenamefont {Zhang},\ and\ \citenamefont
  {Goldman}}]{Wang}%
  \BibitemOpen
  \bibfield  {author} {\bibinfo {author} {\bibfnamefont {T.}~\bibnamefont
  {Wang}}, \bibinfo {author} {\bibfnamefont {K.~M.}\ \bibnamefont {Beauchamp}},
  \bibinfo {author} {\bibfnamefont {D.~D.}\ \bibnamefont {Berkley}}, \bibinfo
  {author} {\bibfnamefont {B.~R.}\ \bibnamefont {Johnson}}, \bibinfo {author}
  {\bibfnamefont {J.-X.}\ \bibnamefont {Liu}}, \bibinfo {author} {\bibfnamefont
  {J.}~\bibnamefont {Zhang}}, \ and\ \bibinfo {author} {\bibfnamefont {A.~M.}\
  \bibnamefont {Goldman}},\ }\bibfield  {title} {\enquote {\bibinfo {title}
  {Onset of high-temperature superconductivity in the two-dimensional limit},}\
  }\href {\doibase 10.1103/PhysRevB.43.8623} {\bibfield  {journal} {\bibinfo
  {journal} {Phys. Rev. B}\ }\textbf {\bibinfo {volume} {43}},\ \bibinfo
  {pages} {8623--8626} (\bibinfo {year} {1991})}\BibitemShut {NoStop}%
\bibitem [{\citenamefont {{Yang}}\ \emph {et~al.}(2022)\citenamefont {{Yang}},
  \citenamefont {{Liu}}, \citenamefont {{Liu}}, \citenamefont {{Wang}},
  \citenamefont {{Qiu}}, \citenamefont {{Wang}}, \citenamefont {{Wang}},
  \citenamefont {{He}}, \citenamefont {{Li}}, \citenamefont {{Li}},
  \citenamefont {{Tang}}, \citenamefont {{Wang}}, \citenamefont {{Xie}},
  \citenamefont {{Valles}}, \citenamefont {{Xiong}},\ and\ \citenamefont
  {{Li}}}]{2022Natur.601..205Y}%
  \BibitemOpen
  \bibfield  {author} {\bibinfo {author} {\bibfnamefont {Chao}\ \bibnamefont
  {{Yang}}}, \bibinfo {author} {\bibfnamefont {Haiwen}\ \bibnamefont {{Liu}}},
  \bibinfo {author} {\bibfnamefont {Yi}~\bibnamefont {{Liu}}}, \bibinfo
  {author} {\bibfnamefont {Jiandong}\ \bibnamefont {{Wang}}}, \bibinfo {author}
  {\bibfnamefont {Dong}\ \bibnamefont {{Qiu}}}, \bibinfo {author}
  {\bibfnamefont {Sishuang}\ \bibnamefont {{Wang}}}, \bibinfo {author}
  {\bibfnamefont {Yang}\ \bibnamefont {{Wang}}}, \bibinfo {author}
  {\bibfnamefont {Qianmei}\ \bibnamefont {{He}}}, \bibinfo {author}
  {\bibfnamefont {Xiuli}\ \bibnamefont {{Li}}}, \bibinfo {author}
  {\bibfnamefont {Peng}\ \bibnamefont {{Li}}}, \bibinfo {author} {\bibfnamefont
  {Yue}\ \bibnamefont {{Tang}}}, \bibinfo {author} {\bibfnamefont {Jian}\
  \bibnamefont {{Wang}}}, \bibinfo {author} {\bibfnamefont {X.~C.}\
  \bibnamefont {{Xie}}}, \bibinfo {author} {\bibfnamefont {James~M.}\
  \bibnamefont {{Valles}}}, \bibinfo {author} {\bibfnamefont {Jie}\
  \bibnamefont {{Xiong}}}, \ and\ \bibinfo {author} {\bibfnamefont {Yanrong}\
  \bibnamefont {{Li}}},\ }\bibfield  {title} {\enquote {\bibinfo {title}
  {{Signatures of a strange metal in a bosonic system}},}\ }\href {\doibase
  10.1038/s41586-021-04239-y} {\bibfield  {journal} {\bibinfo  {journal}
  {\nat}\ }\textbf {\bibinfo {volume} {601}},\ \bibinfo {pages} {205--210}
  (\bibinfo {year} {2022})},\ \Eprint {http://arxiv.org/abs/2105.02654}
  {arXiv:2105.02654 [cond-mat.supr-con]} \BibitemShut {NoStop}%
\bibitem [{\citenamefont {Herbut}(2000)}]{Herbut}%
  \BibitemOpen
  \bibfield  {author} {\bibinfo {author} {\bibfnamefont {Igor~F.}\ \bibnamefont
  {Herbut}},\ }\bibfield  {title} {\enquote {\bibinfo {title} {Zero-temperature
  $\mathit{d}$-wave superconducting phase transition},}\ }\href {\doibase
  10.1103/PhysRevLett.85.1532} {\bibfield  {journal} {\bibinfo  {journal}
  {Phys. Rev. Lett.}\ }\textbf {\bibinfo {volume} {85}},\ \bibinfo {pages}
  {1532--1535} (\bibinfo {year} {2000})}\BibitemShut {NoStop}%
\bibitem [{\citenamefont {Podolsky}\ \emph {et~al.}(2007)\citenamefont
  {Podolsky}, \citenamefont {Vishwanath}, \citenamefont {Moore},\ and\
  \citenamefont {Sachdev}}]{Podolsky}%
  \BibitemOpen
  \bibfield  {author} {\bibinfo {author} {\bibfnamefont {Daniel}\ \bibnamefont
  {Podolsky}}, \bibinfo {author} {\bibfnamefont {Ashvin}\ \bibnamefont
  {Vishwanath}}, \bibinfo {author} {\bibfnamefont {Joel}\ \bibnamefont
  {Moore}}, \ and\ \bibinfo {author} {\bibfnamefont {Subir}\ \bibnamefont
  {Sachdev}},\ }\bibfield  {title} {\enquote {\bibinfo {title} {Thermoelectric
  transport near pair breaking quantum phase transition out of $d$-wave
  superconductivity},}\ }\href {\doibase 10.1103/PhysRevB.75.014520} {\bibfield
   {journal} {\bibinfo  {journal} {Phys. Rev. B}\ }\textbf {\bibinfo {volume}
  {75}},\ \bibinfo {pages} {014520} (\bibinfo {year} {2007})}\BibitemShut
  {NoStop}%
\bibitem [{\citenamefont {{Li}}\ \emph
  {et~al.}(2021{\natexlab{a}})\citenamefont {{Li}}, \citenamefont
  {{Kivelson}},\ and\ \citenamefont {{Lee}}}]{dunghai}%
  \BibitemOpen
  \bibfield  {author} {\bibinfo {author} {\bibfnamefont {Zi-Xiang}\
  \bibnamefont {{Li}}}, \bibinfo {author} {\bibfnamefont {Steven~A.}\
  \bibnamefont {{Kivelson}}}, \ and\ \bibinfo {author} {\bibfnamefont
  {Dung-Hai}\ \bibnamefont {{Lee}}},\ }\bibfield  {title} {\enquote {\bibinfo
  {title} {{Superconductor-to-metal transition in overdoped cuprates}},}\
  }\href {\doibase 10.1038/s41535-021-00335-4} {\bibfield  {journal} {\bibinfo
  {journal} {npj Quantum Materials}\ }\textbf {\bibinfo {volume} {6}},\
  \bibinfo {eid} {36} (\bibinfo {year} {2021}{\natexlab{a}})},\ \Eprint
  {http://arxiv.org/abs/2010.06091} {arXiv:2010.06091 [cond-mat.supr-con]}
  \BibitemShut {NoStop}%
\bibitem [{\citenamefont {Kim}\ \emph {et~al.}(2021)\citenamefont {Kim},
  \citenamefont {Altman},\ and\ \citenamefont {Cao}}]{Kim:2020jpz}%
  \BibitemOpen
  \bibfield  {author} {\bibinfo {author} {\bibfnamefont {Jaewon}\ \bibnamefont
  {Kim}}, \bibinfo {author} {\bibfnamefont {Ehud}\ \bibnamefont {Altman}}, \
  and\ \bibinfo {author} {\bibfnamefont {Xiangyu}\ \bibnamefont {Cao}},\
  }\bibfield  {title} {\enquote {\bibinfo {title} {{Dirac Fast Scramblers}},}\
  }\href {\doibase 10.1103/PhysRevB.103.L081113} {\bibfield  {journal}
  {\bibinfo  {journal} {Phys. Rev. B}\ }\textbf {\bibinfo {volume} {103}},\
  \bibinfo {pages} {081113} (\bibinfo {year} {2021})},\ \Eprint
  {http://arxiv.org/abs/2010.10545} {arXiv:2010.10545 [cond-mat.str-el]}
  \BibitemShut {NoStop}%
\bibitem [{\citenamefont {Aldape}\ \emph {et~al.}(2022)\citenamefont {Aldape},
  \citenamefont {Cookmeyer}, \citenamefont {Patel},\ and\ \citenamefont
  {Altman}}]{Aldape:2020enq}%
  \BibitemOpen
  \bibfield  {author} {\bibinfo {author} {\bibfnamefont {Erik~E.}\ \bibnamefont
  {Aldape}}, \bibinfo {author} {\bibfnamefont {Tessa}\ \bibnamefont
  {Cookmeyer}}, \bibinfo {author} {\bibfnamefont {Aavishkar~A.}\ \bibnamefont
  {Patel}}, \ and\ \bibinfo {author} {\bibfnamefont {Ehud}\ \bibnamefont
  {Altman}},\ }\bibfield  {title} {\enquote {\bibinfo {title} {{Solvable theory
  of a strange metal at the breakdown of a heavy Fermi liquid}},}\ }\href
  {\doibase 10.1103/PhysRevB.105.235111} {\bibfield  {journal} {\bibinfo
  {journal} {Phys. Rev. B}\ }\textbf {\bibinfo {volume} {105}},\ \bibinfo
  {pages} {235111} (\bibinfo {year} {2022})},\ \Eprint
  {http://arxiv.org/abs/2012.00763} {arXiv:2012.00763 [cond-mat.str-el]}
  \BibitemShut {NoStop}%
\bibitem [{\citenamefont {Esterlis}\ \emph {et~al.}(2021)\citenamefont
  {Esterlis}, \citenamefont {Guo}, \citenamefont {Patel},\ and\ \citenamefont
  {Sachdev}}]{esterlis2021large}%
  \BibitemOpen
  \bibfield  {author} {\bibinfo {author} {\bibfnamefont {Ilya}\ \bibnamefont
  {Esterlis}}, \bibinfo {author} {\bibfnamefont {Haoyu}\ \bibnamefont {Guo}},
  \bibinfo {author} {\bibfnamefont {Aavishkar~A}\ \bibnamefont {Patel}}, \ and\
  \bibinfo {author} {\bibfnamefont {Subir}\ \bibnamefont {Sachdev}},\
  }\bibfield  {title} {\enquote {\bibinfo {title} {Large-n theory of critical
  fermi surfaces},}\ }\href@noop {} {\bibfield  {journal} {\bibinfo  {journal}
  {Physical Review B}\ }\textbf {\bibinfo {volume} {103}},\ \bibinfo {pages}
  {235129} (\bibinfo {year} {2021})}\BibitemShut {NoStop}%
\bibitem [{\citenamefont {Patel}\ \emph {et~al.}(2022)\citenamefont {Patel},
  \citenamefont {Guo}, \citenamefont {Esterlis},\ and\ \citenamefont
  {Sachdev}}]{patel2}%
  \BibitemOpen
  \bibfield  {author} {\bibinfo {author} {\bibfnamefont {Aavishkar~A.}\
  \bibnamefont {Patel}}, \bibinfo {author} {\bibfnamefont {Haoyu}\ \bibnamefont
  {Guo}}, \bibinfo {author} {\bibfnamefont {Ilya}\ \bibnamefont {Esterlis}}, \
  and\ \bibinfo {author} {\bibfnamefont {Subir}\ \bibnamefont {Sachdev}},\
  }\bibfield  {title} {\enquote {\bibinfo {title} {{Universal, low temperature,
  $T$-linear resistivity in two-dimensional quantum-critical metals from
  spatially random interactions}},}\ }\href@noop {} {\  (\bibinfo {year}
  {2022})},\ \Eprint {http://arxiv.org/abs/2203.04990} {arXiv:2203.04990
  [cond-mat.str-el]} \BibitemShut {NoStop}%
\bibitem [{\citenamefont {{Giraldo-Gallo}}\ \emph {et~al.}(2018)\citenamefont
  {{Giraldo-Gallo}}, \citenamefont {{Galvis}}, \citenamefont {{Stegen}},
  \citenamefont {{Modic}}, \citenamefont {{Balakirev}}, \citenamefont
  {{Betts}}, \citenamefont {{Lian}}, \citenamefont {{Moir}}, \citenamefont
  {{Riggs}}, \citenamefont {{Wu}}, \citenamefont {{Bollinger}}, \citenamefont
  {{He}}, \citenamefont {{Bozovi{\'c}}}, \citenamefont {{Ramshaw}},
  \citenamefont {{McDonald}}, \citenamefont {{Boebinger}},\ and\ \citenamefont
  {{Shekhter}}}]{LSCO}%
  \BibitemOpen
  \bibfield  {author} {\bibinfo {author} {\bibfnamefont {P.}~\bibnamefont
  {{Giraldo-Gallo}}}, \bibinfo {author} {\bibfnamefont {J.~A.}\ \bibnamefont
  {{Galvis}}}, \bibinfo {author} {\bibfnamefont {Z.}~\bibnamefont {{Stegen}}},
  \bibinfo {author} {\bibfnamefont {K.~A.}\ \bibnamefont {{Modic}}}, \bibinfo
  {author} {\bibfnamefont {F.~F.}\ \bibnamefont {{Balakirev}}}, \bibinfo
  {author} {\bibfnamefont {J.~B.}\ \bibnamefont {{Betts}}}, \bibinfo {author}
  {\bibfnamefont {X.}~\bibnamefont {{Lian}}}, \bibinfo {author} {\bibfnamefont
  {C.}~\bibnamefont {{Moir}}}, \bibinfo {author} {\bibfnamefont {S.~C.}\
  \bibnamefont {{Riggs}}}, \bibinfo {author} {\bibfnamefont {J.}~\bibnamefont
  {{Wu}}}, \bibinfo {author} {\bibfnamefont {A.~T.}\ \bibnamefont
  {{Bollinger}}}, \bibinfo {author} {\bibfnamefont {X.}~\bibnamefont {{He}}},
  \bibinfo {author} {\bibfnamefont {I.}~\bibnamefont {{Bozovi{\'c}}}}, \bibinfo
  {author} {\bibfnamefont {B.~J.}\ \bibnamefont {{Ramshaw}}}, \bibinfo {author}
  {\bibfnamefont {R.~D.}\ \bibnamefont {{McDonald}}}, \bibinfo {author}
  {\bibfnamefont {G.~S.}\ \bibnamefont {{Boebinger}}}, \ and\ \bibinfo {author}
  {\bibfnamefont {A.}~\bibnamefont {{Shekhter}}},\ }\bibfield  {title}
  {\enquote {\bibinfo {title} {{Scale-invariant magnetoresistance in a cuprate
  superconductor}},}\ }\href {\doibase 10.1126/science.aan3178} {\bibfield
  {journal} {\bibinfo  {journal} {Science}\ }\textbf {\bibinfo {volume}
  {361}},\ \bibinfo {pages} {479--481} (\bibinfo {year} {2018})},\ \Eprint
  {http://arxiv.org/abs/1705.05806} {arXiv:1705.05806 [cond-mat.str-el]}
  \BibitemShut {NoStop}%
\bibitem [{\citenamefont {{Cooper}}\ \emph {et~al.}(2009)\citenamefont
  {{Cooper}}, \citenamefont {{Wang}}, \citenamefont {{Vignolle}}, \citenamefont
  {{Lipscombe}}, \citenamefont {{Hayden}}, \citenamefont {{Tanabe}},
  \citenamefont {{Adachi}}, \citenamefont {{Koike}}, \citenamefont {{Nohara}},
  \citenamefont {{Takagi}}, \citenamefont {{Proust}},\ and\ \citenamefont
  {{Hussey}}}]{2009Sci...323..603C}%
  \BibitemOpen
  \bibfield  {author} {\bibinfo {author} {\bibfnamefont {R.~A.}\ \bibnamefont
  {{Cooper}}}, \bibinfo {author} {\bibfnamefont {Y.}~\bibnamefont {{Wang}}},
  \bibinfo {author} {\bibfnamefont {B.}~\bibnamefont {{Vignolle}}}, \bibinfo
  {author} {\bibfnamefont {O.~J.}\ \bibnamefont {{Lipscombe}}}, \bibinfo
  {author} {\bibfnamefont {S.~M.}\ \bibnamefont {{Hayden}}}, \bibinfo {author}
  {\bibfnamefont {Y.}~\bibnamefont {{Tanabe}}}, \bibinfo {author}
  {\bibfnamefont {T.}~\bibnamefont {{Adachi}}}, \bibinfo {author}
  {\bibfnamefont {Y.}~\bibnamefont {{Koike}}}, \bibinfo {author} {\bibfnamefont
  {M.}~\bibnamefont {{Nohara}}}, \bibinfo {author} {\bibfnamefont
  {H.}~\bibnamefont {{Takagi}}}, \bibinfo {author} {\bibfnamefont {Cyril}\
  \bibnamefont {{Proust}}}, \ and\ \bibinfo {author} {\bibfnamefont {N.~E.}\
  \bibnamefont {{Hussey}}},\ }\bibfield  {title} {\enquote {\bibinfo {title}
  {{Anomalous Criticality in the Electrical Resistivity of
  La$_{2-}$$_{x}$Sr$_{x}$CuO$_{4}$}},}\ }\href {\doibase
  10.1126/science.1165015} {\bibfield  {journal} {\bibinfo  {journal}
  {Science}\ }\textbf {\bibinfo {volume} {323}},\ \bibinfo {pages} {603}
  (\bibinfo {year} {2009})}\BibitemShut {NoStop}%
\bibitem [{\citenamefont {{Hayes}}\ \emph {et~al.}(2021)\citenamefont
  {{Hayes}}, \citenamefont {{Maksimovic}}, \citenamefont {{Lopez}},
  \citenamefont {{Chan}}, \citenamefont {{Ramshaw}}, \citenamefont
  {{McDonald}},\ and\ \citenamefont {{Analytis}}}]{Hayes}%
  \BibitemOpen
  \bibfield  {author} {\bibinfo {author} {\bibfnamefont {Ian~M.}\ \bibnamefont
  {{Hayes}}}, \bibinfo {author} {\bibfnamefont {Nikola}\ \bibnamefont
  {{Maksimovic}}}, \bibinfo {author} {\bibfnamefont {Gilbert~N.}\ \bibnamefont
  {{Lopez}}}, \bibinfo {author} {\bibfnamefont {Mun~K.}\ \bibnamefont
  {{Chan}}}, \bibinfo {author} {\bibfnamefont {B.~J.}\ \bibnamefont
  {{Ramshaw}}}, \bibinfo {author} {\bibfnamefont {Ross~D.}\ \bibnamefont
  {{McDonald}}}, \ and\ \bibinfo {author} {\bibfnamefont {James~G.}\
  \bibnamefont {{Analytis}}},\ }\bibfield  {title} {\enquote {\bibinfo {title}
  {{Superconductivity and quantum criticality linked by the Hall effect in a
  strange metal}},}\ }\href {\doibase 10.1038/s41567-020-0982-x} {\bibfield
  {journal} {\bibinfo  {journal} {Nature Physics}\ }\textbf {\bibinfo {volume}
  {17}},\ \bibinfo {pages} {58--62} (\bibinfo {year} {2021})},\ \Eprint
  {http://arxiv.org/abs/1912.06130} {arXiv:1912.06130 [cond-mat.str-el]}
  \BibitemShut {NoStop}%
\bibitem [{\citenamefont {{Ataei}}\ \emph {et~al.}(2022)\citenamefont
  {{Ataei}}, \citenamefont {{Gourgout}}, \citenamefont {{Grissonnanche}},
  \citenamefont {{Chen}}, \citenamefont {{Baglo}}, \citenamefont {{Boulanger}},
  \citenamefont {{Lalibert{\'e}}}, \citenamefont {{Badoux}}, \citenamefont
  {{Doiron-Leyraud}}, \citenamefont {{Oliviero}}, \citenamefont {{Benhabib}},
  \citenamefont {{Vignolles}}, \citenamefont {{Zhou}}, \citenamefont {{Ono}},
  \citenamefont {{Takagi}}, \citenamefont {{Proust}},\ and\ \citenamefont
  {{Taillefer}}}]{2022NatPh..18.1420A}%
  \BibitemOpen
  \bibfield  {author} {\bibinfo {author} {\bibfnamefont {Amirreza}\
  \bibnamefont {{Ataei}}}, \bibinfo {author} {\bibfnamefont {A.}~\bibnamefont
  {{Gourgout}}}, \bibinfo {author} {\bibfnamefont {G.}~\bibnamefont
  {{Grissonnanche}}}, \bibinfo {author} {\bibfnamefont {L.}~\bibnamefont
  {{Chen}}}, \bibinfo {author} {\bibfnamefont {J.}~\bibnamefont {{Baglo}}},
  \bibinfo {author} {\bibfnamefont {M.~E.}\ \bibnamefont {{Boulanger}}},
  \bibinfo {author} {\bibfnamefont {F.}~\bibnamefont {{Lalibert{\'e}}}},
  \bibinfo {author} {\bibfnamefont {S.}~\bibnamefont {{Badoux}}}, \bibinfo
  {author} {\bibfnamefont {N.}~\bibnamefont {{Doiron-Leyraud}}}, \bibinfo
  {author} {\bibfnamefont {V.}~\bibnamefont {{Oliviero}}}, \bibinfo {author}
  {\bibfnamefont {S.}~\bibnamefont {{Benhabib}}}, \bibinfo {author}
  {\bibfnamefont {D.}~\bibnamefont {{Vignolles}}}, \bibinfo {author}
  {\bibfnamefont {J.~S.}\ \bibnamefont {{Zhou}}}, \bibinfo {author}
  {\bibfnamefont {S.}~\bibnamefont {{Ono}}}, \bibinfo {author} {\bibfnamefont
  {H.}~\bibnamefont {{Takagi}}}, \bibinfo {author} {\bibfnamefont
  {C.}~\bibnamefont {{Proust}}}, \ and\ \bibinfo {author} {\bibfnamefont
  {Louis}\ \bibnamefont {{Taillefer}}},\ }\bibfield  {title} {\enquote
  {\bibinfo {title} {{Electrons with Planckian scattering obey standard orbital
  motion in a magnetic field}},}\ }\href {\doibase 10.1038/s41567-022-01763-0}
  {\bibfield  {journal} {\bibinfo  {journal} {Nature Physics}\ }\textbf
  {\bibinfo {volume} {18}},\ \bibinfo {pages} {1420--1424} (\bibinfo {year}
  {2022})},\ \Eprint {http://arxiv.org/abs/2203.05035} {arXiv:2203.05035
  [cond-mat.str-el]} \BibitemShut {NoStop}%
\bibitem [{\citenamefont {{Jaoui}}\ \emph {et~al.}(2022)\citenamefont
  {{Jaoui}}, \citenamefont {{Das}}, \citenamefont {{Di Battista}},
  \citenamefont {{D{\'\i}ez-M{\'e}rida}}, \citenamefont {{Lu}}, \citenamefont
  {{Watanabe}}, \citenamefont {{Taniguchi}}, \citenamefont {{Ishizuka}},
  \citenamefont {{Levitov}},\ and\ \citenamefont {{Efetov}}}]{Jaoui}%
  \BibitemOpen
  \bibfield  {author} {\bibinfo {author} {\bibfnamefont {Alexandre}\
  \bibnamefont {{Jaoui}}}, \bibinfo {author} {\bibfnamefont {Ipsita}\
  \bibnamefont {{Das}}}, \bibinfo {author} {\bibfnamefont {Giorgio}\
  \bibnamefont {{Di Battista}}}, \bibinfo {author} {\bibfnamefont {Jaime}\
  \bibnamefont {{D{\'\i}ez-M{\'e}rida}}}, \bibinfo {author} {\bibfnamefont
  {Xiaobo}\ \bibnamefont {{Lu}}}, \bibinfo {author} {\bibfnamefont {Kenji}\
  \bibnamefont {{Watanabe}}}, \bibinfo {author} {\bibfnamefont {Takashi}\
  \bibnamefont {{Taniguchi}}}, \bibinfo {author} {\bibfnamefont {Hiroaki}\
  \bibnamefont {{Ishizuka}}}, \bibinfo {author} {\bibfnamefont {Leonid}\
  \bibnamefont {{Levitov}}}, \ and\ \bibinfo {author} {\bibfnamefont
  {Dmitri~K.}\ \bibnamefont {{Efetov}}},\ }\bibfield  {title} {\enquote
  {\bibinfo {title} {{Quantum critical behaviour in magic-angle twisted bilayer
  graphene}},}\ }\href {\doibase 10.1038/s41567-022-01556-5} {\bibfield
  {journal} {\bibinfo  {journal} {Nature Physics}\ }\textbf {\bibinfo {volume}
  {18}},\ \bibinfo {pages} {633--638} (\bibinfo {year} {2022})},\ \Eprint
  {http://arxiv.org/abs/2108.07753} {arXiv:2108.07753 [cond-mat.str-el]}
  \BibitemShut {NoStop}%
\bibitem [{Note1()}]{Note1}%
  \BibitemOpen
  \bibinfo {note} {In principle, the number of bosons and fermions can be
  different as long as they are taken to infinity together (i.e. with a
  constant ratio), but this detail does not greatly affect the physics of our
  model.}\BibitemShut {Stop}%
\bibitem [{\citenamefont {{Varma}}(2016)}]{Chandra}%
  \BibitemOpen
  \bibfield  {author} {\bibinfo {author} {\bibfnamefont {Chandra~M.}\
  \bibnamefont {{Varma}}},\ }\bibfield  {title} {\enquote {\bibinfo {title}
  {{Quantum-critical fluctuations in 2D metals: strange metals and
  superconductivity in antiferromagnets and in cuprates}},}\ }\href {\doibase
  10.1088/0034-4885/79/8/082501} {\bibfield  {journal} {\bibinfo  {journal}
  {Reports on Progress in Physics}\ }\textbf {\bibinfo {volume} {79}},\
  \bibinfo {eid} {082501} (\bibinfo {year} {2016})},\ \Eprint
  {http://arxiv.org/abs/1601.03403} {arXiv:1601.03403 [cond-mat.str-el]}
  \BibitemShut {NoStop}%
\bibitem [{Note2()}]{Note2}%
  \BibitemOpen
  \bibinfo {note} {Although the fermions also contribute to the thermal boson
  gap, this contribution scales as $T^2/\Lambda _f$ and may be
  ignored}\BibitemShut {NoStop}%
\bibitem [{\citenamefont {Mahan}(2000)}]{mah00}%
  \BibitemOpen
  \bibfield  {author} {\bibinfo {author} {\bibfnamefont {G.~D.}\ \bibnamefont
  {Mahan}},\ }\href@noop {} {\emph {\bibinfo {title} {Many Particle Physics,
  Third Edition}}}\ (\bibinfo  {publisher} {Plenum},\ \bibinfo {address} {New
  York},\ \bibinfo {year} {2000})\BibitemShut {NoStop}%
\bibitem [{\citenamefont {{Kim}}\ \emph {et~al.}(1995)\citenamefont {{Kim}},
  \citenamefont {{Lee}},\ and\ \citenamefont {{Wen}}}]{KimLeeWen}%
  \BibitemOpen
  \bibfield  {author} {\bibinfo {author} {\bibfnamefont {Yong~Baek}\
  \bibnamefont {{Kim}}}, \bibinfo {author} {\bibfnamefont {Patrick~A.}\
  \bibnamefont {{Lee}}}, \ and\ \bibinfo {author} {\bibfnamefont {Xiao-Gang}\
  \bibnamefont {{Wen}}},\ }\bibfield  {title} {\enquote {\bibinfo {title}
  {{Quantum Boltzmann equation of composite fermions interacting with a gauge
  field}},}\ }\href {\doibase 10.1103/PhysRevB.52.17275} {\bibfield  {journal}
  {\bibinfo  {journal} {\prb}\ }\textbf {\bibinfo {volume} {52}},\ \bibinfo
  {pages} {17275--17292} (\bibinfo {year} {1995})},\ \Eprint
  {http://arxiv.org/abs/cond-mat/9504063} {arXiv:cond-mat/9504063 [cond-mat]}
  \BibitemShut {NoStop}%
\bibitem [{\citenamefont {{Nave}}\ and\ \citenamefont {{Lee}}(2007)}]{NaveLee}%
  \BibitemOpen
  \bibfield  {author} {\bibinfo {author} {\bibfnamefont {Cody~P.}\ \bibnamefont
  {{Nave}}}\ and\ \bibinfo {author} {\bibfnamefont {Patrick~A.}\ \bibnamefont
  {{Lee}}},\ }\bibfield  {title} {\enquote {\bibinfo {title} {{Transport
  properties of a spinon Fermi surface coupled to a U(1) gauge field}},}\
  }\href {\doibase 10.1103/PhysRevB.76.235124} {\bibfield  {journal} {\bibinfo
  {journal} {\prb}\ }\textbf {\bibinfo {volume} {76}},\ \bibinfo {eid} {235124}
  (\bibinfo {year} {2007})},\ \Eprint {http://arxiv.org/abs/0708.1850}
  {arXiv:0708.1850 [cond-mat.str-el]} \BibitemShut {NoStop}%
\bibitem [{\citenamefont {{Senthil}}\ \emph {et~al.}(2004)\citenamefont
  {{Senthil}}, \citenamefont {{Vojta}},\ and\ \citenamefont
  {{Sachdev}}}]{2004PhRvB..69c5111S}%
  \BibitemOpen
  \bibfield  {author} {\bibinfo {author} {\bibfnamefont {T.}~\bibnamefont
  {{Senthil}}}, \bibinfo {author} {\bibfnamefont {Matthias}\ \bibnamefont
  {{Vojta}}}, \ and\ \bibinfo {author} {\bibfnamefont {Subir}\ \bibnamefont
  {{Sachdev}}},\ }\bibfield  {title} {\enquote {\bibinfo {title} {{Weak
  magnetism and non-Fermi liquids near heavy-fermion critical points}},}\
  }\href {\doibase 10.1103/PhysRevB.69.035111} {\bibfield  {journal} {\bibinfo
  {journal} {\prb}\ }\textbf {\bibinfo {volume} {69}},\ \bibinfo {eid} {035111}
  (\bibinfo {year} {2004})},\ \Eprint {http://arxiv.org/abs/cond-mat/0305193}
  {arXiv:cond-mat/0305193 [cond-mat.str-el]} \BibitemShut {NoStop}%
\bibitem [{\citenamefont {{Maslov}}\ \emph {et~al.}(2011)\citenamefont
  {{Maslov}}, \citenamefont {{Yudson}},\ and\ \citenamefont
  {{Chubukov}}}]{Maslov}%
  \BibitemOpen
  \bibfield  {author} {\bibinfo {author} {\bibfnamefont {Dmitrii~L.}\
  \bibnamefont {{Maslov}}}, \bibinfo {author} {\bibfnamefont {Vladimir~I.}\
  \bibnamefont {{Yudson}}}, \ and\ \bibinfo {author} {\bibfnamefont
  {Andrey~V.}\ \bibnamefont {{Chubukov}}},\ }\bibfield  {title} {\enquote
  {\bibinfo {title} {{Resistivity of a Non-Galilean-Invariant Fermi Liquid near
  Pomeranchuk Quantum Criticality}},}\ }\href {\doibase
  10.1103/PhysRevLett.106.106403} {\bibfield  {journal} {\bibinfo  {journal}
  {\prl}\ }\textbf {\bibinfo {volume} {106}},\ \bibinfo {eid} {106403}
  (\bibinfo {year} {2011})},\ \Eprint {http://arxiv.org/abs/1012.0069}
  {arXiv:1012.0069 [cond-mat.str-el]} \BibitemShut {NoStop}%
\bibitem [{Note3()}]{Note3}%
  \BibitemOpen
  \bibinfo {note} {$\sigma _f^{xy}$ is identically zero without a magnetic
  field present.}\BibitemShut {Stop}%
\bibitem [{\citenamefont {Doniach}(1981)}]{doniach}%
  \BibitemOpen
  \bibfield  {author} {\bibinfo {author} {\bibfnamefont {S.}~\bibnamefont
  {Doniach}},\ }\bibfield  {title} {\enquote {\bibinfo {title} {Quantum
  fluctuations in two-dimensional superconductors},}\ }\href {\doibase
  10.1103/PhysRevB.24.5063} {\bibfield  {journal} {\bibinfo  {journal} {Phys.
  Rev. B}\ }\textbf {\bibinfo {volume} {24}},\ \bibinfo {pages} {5063--5070}
  (\bibinfo {year} {1981})}\BibitemShut {NoStop}%
\bibitem [{\citenamefont {{Li}}\ \emph
  {et~al.}(2021{\natexlab{b}})\citenamefont {{Li}}, \citenamefont {{Liu}},
  \citenamefont {{Liu}}, \citenamefont {{Qi}}, \citenamefont {{Ji}},
  \citenamefont {{Dong}}, \citenamefont {{Sun}}, \citenamefont {{Zhang}},
  \citenamefont {{Ji}}, \citenamefont {{Cui}}, \citenamefont {{Samarth}},
  \citenamefont {{Wang}}, \citenamefont {{Xie}}, \citenamefont {{Xue}},\ and\
  \citenamefont {{Wang}}}]{YananLiPreprint}%
  \BibitemOpen
  \bibfield  {author} {\bibinfo {author} {\bibfnamefont {Yanan}\ \bibnamefont
  {{Li}}}, \bibinfo {author} {\bibfnamefont {Yi}~\bibnamefont {{Liu}}},
  \bibinfo {author} {\bibfnamefont {Haiwen}\ \bibnamefont {{Liu}}}, \bibinfo
  {author} {\bibfnamefont {Shichao}\ \bibnamefont {{Qi}}}, \bibinfo {author}
  {\bibfnamefont {Haoran}\ \bibnamefont {{Ji}}}, \bibinfo {author}
  {\bibfnamefont {Wenfeng}\ \bibnamefont {{Dong}}}, \bibinfo {author}
  {\bibfnamefont {Yi}~\bibnamefont {{Sun}}}, \bibinfo {author} {\bibfnamefont
  {Wenhao}\ \bibnamefont {{Zhang}}}, \bibinfo {author} {\bibfnamefont
  {Chengcheng}\ \bibnamefont {{Ji}}}, \bibinfo {author} {\bibfnamefont {Zihan}\
  \bibnamefont {{Cui}}}, \bibinfo {author} {\bibfnamefont {Nitin}\ \bibnamefont
  {{Samarth}}}, \bibinfo {author} {\bibfnamefont {Lili}\ \bibnamefont
  {{Wang}}}, \bibinfo {author} {\bibfnamefont {X.~C.}\ \bibnamefont {{Xie}}},
  \bibinfo {author} {\bibfnamefont {Qi-Kun}\ \bibnamefont {{Xue}}}, \ and\
  \bibinfo {author} {\bibfnamefont {Jian}\ \bibnamefont {{Wang}}},\ }\bibfield
  {title} {\enquote {\bibinfo {title} {{Bosonic metal states in crystalline
  iron-based superconductors at the two-dimensional limit}},}\ }\href@noop {}
  {\bibfield  {journal} {\bibinfo  {journal} {arXiv e-prints}\ ,\ \bibinfo
  {eid} {arXiv:2111.15488}} (\bibinfo {year} {2021}{\natexlab{b}})},\ \Eprint
  {http://arxiv.org/abs/2111.15488} {arXiv:2111.15488 [cond-mat.supr-con]}
  \BibitemShut {NoStop}%
\bibitem [{\citenamefont {Kapitulnik}\ \emph {et~al.}(2019)\citenamefont
  {Kapitulnik}, \citenamefont {Kivelson},\ and\ \citenamefont
  {Spivak}}]{AMrmp}%
  \BibitemOpen
  \bibfield  {author} {\bibinfo {author} {\bibfnamefont {Aharon}\ \bibnamefont
  {Kapitulnik}}, \bibinfo {author} {\bibfnamefont {Steven~A.}\ \bibnamefont
  {Kivelson}}, \ and\ \bibinfo {author} {\bibfnamefont {Boris}\ \bibnamefont
  {Spivak}},\ }\bibfield  {title} {\enquote {\bibinfo {title} {Colloquium:
  Anomalous metals: Failed superconductors},}\ }\href {\doibase
  10.1103/RevModPhys.91.011002} {\bibfield  {journal} {\bibinfo  {journal}
  {Rev. Mod. Phys.}\ }\textbf {\bibinfo {volume} {91}},\ \bibinfo {pages}
  {011002} (\bibinfo {year} {2019})}\BibitemShut {NoStop}%
\bibitem [{\citenamefont {Little}\ and\ \citenamefont
  {Parks}(1962)}]{LittleParks}%
  \BibitemOpen
  \bibfield  {author} {\bibinfo {author} {\bibfnamefont {W.~A.}\ \bibnamefont
  {Little}}\ and\ \bibinfo {author} {\bibfnamefont {R.~D.}\ \bibnamefont
  {Parks}},\ }\bibfield  {title} {\enquote {\bibinfo {title} {Observation of
  quantum periodicity in the transition temperature of a superconducting
  cylinder},}\ }\href {\doibase 10.1103/PhysRevLett.9.9} {\bibfield  {journal}
  {\bibinfo  {journal} {Phys. Rev. Lett.}\ }\textbf {\bibinfo {volume} {9}},\
  \bibinfo {pages} {9--12} (\bibinfo {year} {1962})}\BibitemShut {NoStop}%
\end{thebibliography}%

\newpage
\appendix
\begin{widetext}
\section{Little-Parks Effect} \label{app:LittleParks}
The authors of the experimental paper on nanopatterned YBCO films Ref.~\cite{2022Natur.601..205Y} interpreted $T$ linear metallic conductivity as arising from critical bosonic Cooper pairs. This interpretation was based on the  oscillation of the conductivity as a function of magnetic field with a period corresponding to a bosonic flux quantum $h/2e$ similar to the Little-Parks effect \cite{LittleParks}. 

Here we argue that these oscillations are consistent with our theory wherein the linear in T resistivity comes from the fermions and the bosons add a sub-leading contribution. 
Indirect evidence for this interpretation is given in Fig. \ref{fig:osci}; it shows that the oscillation amplitude $A_\sigma$ shows a logarithmic temperature dependence, matching the predicted scaling of the sub-leading bosonic conductivity. 

\begin{figure}[h]
    \centering
    \includegraphics[width=0.4\columnwidth]{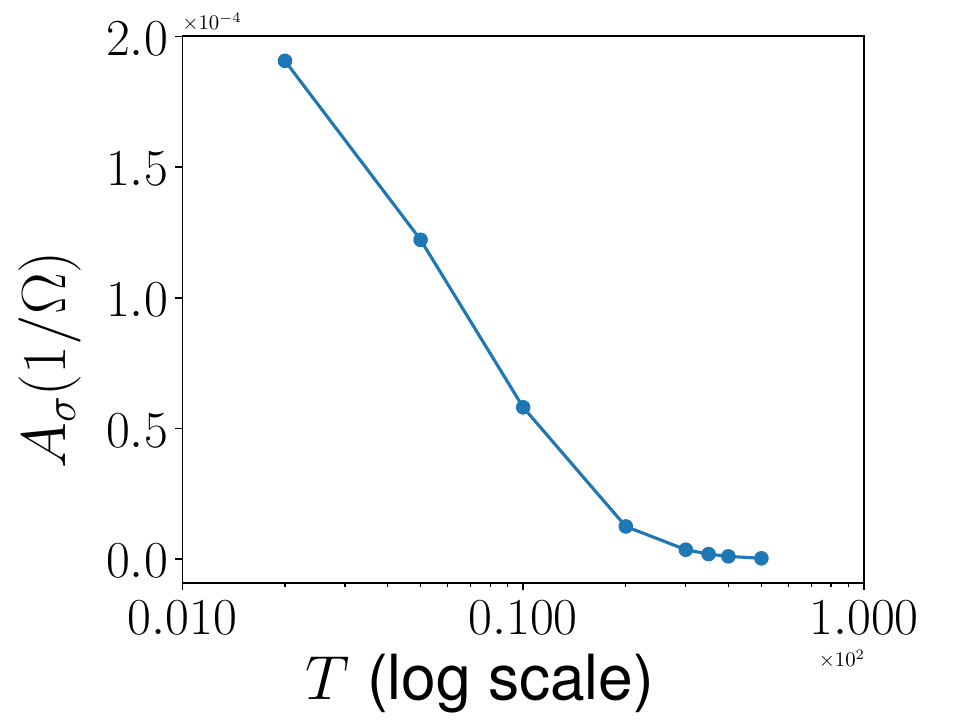}
    \caption{$A_\sigma$ versus $T$ ($\log$ scale) at the critical point for sample $f2$.}
    \label{fig:osci}
\end{figure}

\section{Derivation of the Schwinger-Dyson Equations and Self Energies} \label{app:SD}
In this appendix, we derive the large $N$ Schwinger-Dyson equations and provide the details behind the various self-energy calculations.
In our model, the interaction between the fermions $c_{i,k}^T = (c_{i\uparrow,k }, c_{i\downarrow,k })$ and bosons $\Psi_{n,q}$ is given by,
\begin{equation*}
    H_{int} = \int \frac{d^2k \, d^2q \, d^2q'}{(2\pi)^6} \sum_{ij,n}^N J_k g_{ij,q}^n \Psi_{n,q'}  \left({c}^T_{i \frac{q+q'}{2}+k} \sigma_y c_{j \frac{q+q'}{2}-k} \right)_x + c.c.
\end{equation*}

Averaging over the random disorder $g_{ij,q}^n$ results in the following contribution to the action
\begin{equation}
    S_{int} = \frac{\pi\xi^2}{N^2}\int_{k,q,q',\tau,\tau'}\sum_{ij,n} \Big({c}^T_{i\frac{q+q'}{2}+k} \sigma_y c_{j\frac{q+q'}{2}-k}\Big)_{\tau} 
    \Big({c}^\dagger_{j\frac{q+q'}{2}-k} \sigma_y {c}^*_{i\frac{q+q'}{2}+k} \Big)_{\tau'} J_k^2 e^{-\xi^2 {q'}^2} \Psi^\dagger_n(q,\tau) \Psi_n(q,\tau') \,.
\end{equation}

We now introduce the fermion and boson Green's functions,
$$G(x,\tau,x',\tau') = \frac1{2N} \sum_{i,\sigma} \braket{c_{i\sigma}(x,\tau) c^\dagger_{i\sigma}(x',\tau')}\,, \quad
F(x,\tau,x',\tau') = \frac1N \sum_{n} \braket{\Psi_n(x,\tau) \Psi^\dagger_n(x',\tau')} \,,$$
which we impose through the integral expression for the delta function:
\begin{equation}
\begin{split}
    \delta\Big(\sum_{i\sigma} c_{i\sigma}(x,\tau) c^\dagger_{i\sigma}(x',\tau') - 2N G_{x-x',\tau-\tau'}\Big) =& \int D\Sigma \exp\Big(\Sigma_{x'-x,\tau'-\tau}\big(\sum_{i\sigma} c_{i\sigma}(x,\tau) c^\dagger_{i\sigma}(x',\tau') - 2N G_{x-x',\tau-\tau'}\big)\Big) \\
    \delta\Big(\sum_n \Psi_n(\tau,x) \Psi^\dagger_n(\tau',x') - NF_{\tau-\tau',x-x'} \Big) =& \int D\Pi \exp\Big(\Pi_{\tau'-\tau,x'-x} \big(\sum_n \Psi_n(\tau,x) \Psi_n^\dagger(\tau',x') - NF_{\tau-\tau',x-x'} \big)\Big)
\end{split}
\end{equation}

With the introduction of the Green's functions and their corresponding self energies, the effective action takes the form
\begin{equation}
\begin{split}
    \frac{\beta S(G,\Sigma,F,\Pi)}{N} &= 
    \sum_{\omega_n} \int_k -2 \log \big(-\im\omega_n -\Sigma_{k,\omega_n} + v_F(k-k_F)\big) + \log(\Delta_b + k^2/2m_b - \Pi_{k,\omega_n} )  \\
    & \ \ + 2\Sigma_{k,\omega_n} G_{k,\omega_n} + \Pi_{k,\omega_n} F_{k,\omega_n} - \int_{\tau, \tau', k, q, q'} 2\pi \xi^2  J_k^2 e^{-\xi^2 q^2} \ G_{k+\frac{q+q'}{2},\tau-\tau'} G_{-k+\frac{q+q'}{2},\tau-\tau'}  F_{q,\tau'-\tau}
    %\sum_{\omega} \int_k \big(-\im\omega-\Sigma_{\omega,k}-v_F(k-k_F)\big) \psi^\dagger_{\omega,k}\psi_{\omega,k} + (m_b^2 + k^2 - \Pi_{\omega,k} ) \Psi^\dagger_{\omega,k}  \Psi_{\omega,k} \\
    %& \ \ - \int_{\tau, \tau', k, q, q'} J_k^2 \xi^2 \ G_{k+\frac{q+q'}{2},\tau-\tau'} G_{-k+\frac{q+q'}{2},\tau-\tau'}  F_{q,\tau'-\tau} + \Sigma_{\tau'-\tau,x'-x} G_{\tau-\tau',x-x'} + \gamma \Pi_{\tau'-\tau,x'-x} F_{\tau-\tau',x-x'}
\end{split}
\end{equation}
The saddle point of this action gives the Schwinger-Dyson equations \eqref{eq:dSDeq}.

We now calculate the self energies in \eqref{eq:dSDeq}. We begin with the boson self-energy $\Pi$ \eqref{eq:bosonPid}, which is given by the following integral
\begin{equation}
\begin{split}
    \Pi(q,\Omega) &= \int \frac{k dk d\theta_k}{(2\pi^2)} \int_{\mathcal{B}\left(\frac\pi\xi\right)} \frac{d^2q'}{(2\pi)^2} \int \frac{d\omega}{2\pi} 2\xi^2 J_k^2 \ G(k+(q+q')/2,\omega+\Omega/2) G(-k+(q+q')/2,- \omega+\Omega/2) \\
    &= \int \frac{k dk d\theta_k}{(2\pi^2)} \int_{\mathcal{B}\left(\frac\pi\xi\right)} \frac{d^2q'}{(2\pi)^2} \int \frac{d\omega}{2\pi} 2\xi^2 J_k^2 \ \frac{ G_{-k+(q+q')/2, -\omega+\Omega/2} - G_{k+(q+q')/2,\omega+\Omega/2}}{2\im\omega + \Sigma_{k+(q+q')/2,\omega+\Omega/2} - \Sigma_{-k+(q+q')/2,-\omega+\Omega/2} + v_F (q+q') \cos\theta_{qk}} \,.
\end{split}
\end{equation}
In the second line, we assumed a circular Fermi surface for simplicity.
We also approximate $J_k \sim J \cos 2\theta_k$.

To see the general structure of the solution we first calculate this using the free Green's functions of the fermions (i.e. neglecting the fermion self energy correction).
At this point we also linearize the dispersion about the Fermi energy and neglect the Fermi surface curvature, thereby imposing an effective particle hole symmetry.
Later we discuss the particle hole symmetry breaking terms.
With the above approximations the boson self energy is given by the integral,
\begin{equation}
\begin{split}
    \Pi(q, \Omega) &= \frac{2J^2 \xi^2 k_F}{(2\pi)^5v_F} \int d\theta_k \cos^2 2\theta_k \int_{\mathcal{B}\left(\frac\pi\xi\right)} d^2q' \int d\omega \frac{\im\pi \big(\sgn(\omega+\Omega/2)+\sgn(\omega-\Omega/2)\big)}{2\im\omega + v_F (q+q') \cos\theta_{q+q',k}} \\
    &= \frac{2\pi^2 J^2 \xi^2 k_F}{(2\pi)^5 v_F} \int d\omega  \int_{\mathcal{B}\left(\frac\pi\xi\right)} d^2q' \frac{\sgn(\omega)\big(\sgn(\omega+\Omega/2)+\sgn(\omega-\Omega/2)\big)}{\Big(4\omega^2  + v_F^2 (q+q')^2 \Big)^{1/2}}
    \label{eq:appPi}
\end{split}
\end{equation}
Now we can expand this for small frequency $\Omega$ and momentum $q$. Setting $\Omega = 0$, the momentum dependence of $\Pi$ is given by
\begin{equation}
\begin{split}
    \Pi(q,0) &= \frac{4\pi^2 J^2 \xi^2 k_F}{(2\pi)^5 v_F} \int d\omega \int_{\mathcal{B}\left(\frac\pi\xi\right)} d^2q' (4\omega^2 + v_F^2(q+q')^2)^{-1/2} \simeq \Pi(0,0) + \frac{q^2}{2M}  \\
    \frac1M &\simeq \nabla_q^2 \Pi |_{q=0, \omega =0} = \frac{4\pi^2 J^2 \xi^2 v_F k_F}{(2\pi)^5} \int_\omega \int_{\mathcal{B}\left(\frac\pi\xi\right)} \frac{-8\omega^2 + v_F^2 q'^2}{(4\omega^2 + v_F^2 q'^2)^{5/2}} 
    \simeq -\frac{J^2 \xi^2 k_F}{8\pi^2 v_F}
    \label{eq:appPi_q}
\end{split}
\end{equation}
This result defines the mass renormalization of the boson from the bare mass $m_b$ to $M_b = \Big({M^{-1}+m_b^{-1}}\Big)^{-1}$.
%We see that at small momenta $q \ll \frac{1}{\xi}$ the effective mass of the boson is strongly renormalized becoming of order $1/\xi^2$.

The frequency dependence of the self energy is found by setting $q=0$ 
\begin{equation}
\begin{split}
    \Pi(0) - \Pi(0,\Omega) &= \frac{4\pi^2 J^2 k_F}{(2\pi)^5 v_F} \int_{-\Omega/2}^{\Omega/2} d\omega \int^{\pi/\xi}_0 2\pi q dq \Big(4\omega^2  + v_F^2 q^2 \Big)^{-1/2} \\
    &\simeq \frac{J^2 k_F \xi |\Omega|}{4\pi v_F^2} \,.
    \label{eq:appPi_omega}
\end{split}
\end{equation}
This is reminiscent of the familiar Landau damped form of the self energy $\Delta\Pi(q,\Omega)\propto |\Omega|/q$ with the disorder scale $\xi$ playing the role of $1/q$.

The correction to $\Pi$ from breaking of particle hole symmetry is given by 
\begin{equation}
    \begin{split}
        \Delta \Pi(\Omega) &= \frac{2\pi J^2 \xi^2}{(2\pi)^5} \int \Delta k d\Delta k \int q dq d\theta_q \int d\omega \frac{G(k+q/2,\omega+\Omega/2) - G(-k+q/2,-\omega+\Omega/2)}{2\im\omega + v_F q \cos \theta_{qk}} \\
        & - \frac{2\pi J^2 \xi^2 k_F}{(2\pi)^5} \int d\Delta k \int q dq d\theta_q \int d\omega \frac{G(k+q/2,\omega+\Omega/2) - G(-k+q/2,-\omega+\Omega/2)}{(2\im\omega + v_F q \cos \theta_{qk})^2 } \frac{q \Delta k \cos \theta_{qk}}{m}.
    \end{split}
\end{equation}
%We now derive the particle-hole symmetry breaking term $\eta$ appearing in \eqref{eq:bosonPid} 
Here $\Delta k = |k|-k_F$ denotes the distance from the Fermi surface. The first term above accounts for the non vanishing  curvature of the Fermi surface (i.e. the correction to approximating $k dk$ as $k_F dk$ in \eqref{eq:appPi}).
The second term stems from the non-linear correction to the dispersion near the Fermi surface.
Computing the integral  we find,
\begin{equation}
    \begin{split}
        \Delta \Pi(\Omega)% &= -\frac{2\im \pi J^2 \xi^2}{(2\pi)^4} \int \Delta k d\Delta k \int q dq \int \sgn(\omega) d\omega \frac{G(k+q/2,\omega+\Omega/2) - G(-k+q/2,-\omega+\Omega/2)}{(4\omega^2 + v_F^2 q^2)^{1/2}} \\
        % & - \frac{2\im \pi J^2 \xi^2 k_F v_F}{(2\pi)^4 m} \int \Delta k d\Delta k \int q^3 dq d\theta_q \int \sgn(\omega) d\omega \frac{G(k+q/2,\omega+\Omega/2) - G(-k+q/2,-\omega+\Omega/2)}{ (4\omega^2 + v_F^2 q^2)^{3/2} } \\
        % &= \frac{\im \pi^2 J^2 \xi^2}{4\pi^2 v_F^2} \int q dq \int \sgn(\omega) d\omega \Big(|\omega+\Omega/2|-|-\omega+\Omega/2|) \Big) \frac{2\omega^2 + v_F^2 q^2}{(4\omega^2 + v_F^2 q^2)^{3/2}} \\
        % &= \frac{\im \pi^2 J^2 \xi^2}{8\pi^2 v_F^4} \int \sgn(\omega) d\omega \left(\frac{6\omega^2 + v_F^2/\xi^2}{\sqrt{4\omega^2 + v_F^2/\xi^2}}-3|\omega|\right) \Big(|\omega+\Omega/2|-|-\omega+\Omega/2|) \Big) \\
        &\simeq \im \left(\frac{J^2}{v_F^2} \log \frac{\Lambda}{v_F/\xi} \right) \Omega
    \end{split}
\end{equation}
which gives the value of $\eta$  appearing in \eqref{eq:bosonPid}. 
%Again, although we have treated the fermions as free fermions and ignored the fermion self-energy, this above result does not greatly change in the presence of interactions.

We now turn to the fermions and calculate their self-energy at zero temperature. This will then allow us to show that the above form of $\Pi$ is self consistent and not modified except by a possible prefactor.
The fermion self energy at  momentum $k$ on the fermi surface is given by
%However, the structure of the Landau damping and the mass renormalization of the boson remains even when we include the fermion self energy in the calculation. We will justify this 
\begin{comment}
Going from the first to the second equality, we have used the fact that
\begin{equation}
    \int d \epsilon \frac{\epsilon}{-\im\omega - \Sigma_\omega -\epsilon} = \int d \epsilon \left(-1 + \frac{\im\omega + \Sigma_\omega}{-\im\omega - \Sigma_\omega - \epsilon } \right) = -\Lambda + \im\pi \sgn(\omega) (\im\omega+ \Sigma_\omega)
\end{equation}
\end{comment}
\begin{equation}
\begin{split}
    \Sigma(k,\Omega) &= 2J^2 \xi^2  \int \frac{d^2 q d^2 q' d\omega}{(2\pi)^5} \cos^2 2\theta_{k-\frac{q'}2} F(q,\omega+\Omega) G(-k+q', \omega) \\
    &= \frac{J^2 \xi^2 M_b}{\pi} \int \frac{dq'_{\perp} dq'_{\parallel} d\omega}{(2\pi)^3} \cos^2 2\theta_{k-\frac{q'}2} \log \frac{\Lambda}{\alpha|\omega+\Omega|} \frac{1}{-\im\omega-\Sigma_{-k+q',\omega}-v_F q'_{\perp}} \\
    &= \frac{\im J^2 \xi^2 M_b}{2\pi v_F} \int \frac{dq'_{\parallel} d\omega}{(2\pi)^2} \cos^2 2\theta_{k-\frac{q'}2} \log \frac{\Lambda}{\alpha|\omega+\Omega|} \ \sgn(\omega) \\
    &= \im \gamma_k \Omega \log \frac{\Lambda}{\Omega}\,,
    \label{eq:appSigmaE}
\end{split}
\end{equation}
where,
\begin{equation*}
\begin{split}
    \gamma_k %&=  \frac{2J^2 \xi^2 M_b}{(2\pi)^3 v_F} \int dq'_{\parallel} \cos^2 2\theta_{k-q'/2} 
    &= \frac{J^2 \xi^2 M_b}{4\pi^3 v_F} \int_{-\pi/\xi}^{\pi/\xi} dq'_{\parallel} \cos^2 (2\theta_k - q'_{\parallel}/k_F) \\
    & = \frac{J^2 \xi^2 M_b}{4\pi^3 v_F} \left\{\frac\pi{\xi} + \frac{k_F \cos 4\theta_k \sin(2\pi/k_F \xi)}{2} \right\} \\
    & \simeq \frac{2}{\pi} \left\{\frac{2\pi}{k_F \xi} + \cos 4\theta_k \sin(2\pi/k_F \xi) \right\}
\end{split}
\end{equation*}
We see that the fermion self-energy scales with frequency as $\Omega \log \Lambda/\Omega$.
Importantly, it has a prefactor $\gamma_k$ which is at most $\BigO(1/k_F\xi)$. 
This means that the fermion self-energy $\Sigma$ is parametrically smaller than $\omega$ in the limit of large $k_F\xi$.
Ergo, including the fermion self-energy in \eqref{eq:appPi} results in only a small correction that can be neglected for large values of $k_F\xi$.

We now determine the self-energies at finite temperatures.
The effect of the temperature on the boson self energy is to generate a thermal gap $\Delta_b$ (i.e. a frequency and momentum independent contribution to the self energy)    \cite{Podolsky}.
This thermal gap in turn leads to  a modification of the fermion self-energy \eqref{eq:appSigmaE} to,
\begin{equation}
\begin{split}
    \Sigma(k,\Omega) &= \im\gamma_k \int \frac{\omega}{2\pi} \log \frac{\Lambda}{\Delta_b + \alpha|\omega+\Omega|} \sgn(\omega) \\
    &\simeq \im\gamma_k \log \frac{\Lambda}{\Delta_b} \Omega \,, \textrm{ for } \Omega \ll \Delta_b.
    \label{eq:appSigmaE_finiteT}
\end{split}
\end{equation}
Analytically continuing \eqref{eq:appSigmaE_finiteT} to real time, we find the real part of the retarded fermion self energy $\Re\{\Sigma_R\} \simeq \gamma_k \log \frac{\Lambda}{\Delta_b} \Omega \,,$.
This subsequently results in a renormalization of the quasi-particle weight to \eqref{eq:Z} of the main text.

We now calculate the imaginary part of the retarded fermion self-energy, which (together with the quasi-particle weight $Z$) is related to the quasi-particle decay rate.
Using the standard Keldysh formalism, the retarded fermion self-energy is given as,
\begin{equation*}
    \begin{split}
        \Sigma_R(k,t) &= J^2 \xi^2 \int \frac{d^2q d^2q'}{(2\pi)} \cos^2 2\theta_{k-\frac{q+q'}{2}} \Big(F_R(q,t) G_K(-k+q',-t) - F_K(q,t) G_R(-k+q',-t) \Big) \\
        & \textrm{ where } F_R(t) = -\im\theta(t)\big(F_>(t) - F_<(t)\big) \,, \ \ F_K(t) = F_>(t) + F_<(t) \,, \\
        & \quad G_R(t) = -\im\theta(t)\big(G_>(t) + G_<(t)\big) \,, \ \ G_K(t) = G_>(t) - G_<(t) \,.
    \end{split}
\end{equation*}
where the greater and lesser Green's functions are defined as, $$G_>(t) = \braket{\psi(t)\psi^\dagger(0)}, \ G_<(t) = \braket{\psi^\dagger(0)\psi(t)}\,, \ F_>(t) = \braket{\Psi(t)\Psi^\dagger(0)}, \ F_<(t) = \braket{\Psi^\dagger(0)\Psi(t)}\,.$$

Using the fluctuation-dissipation theorem, the Keldysh Green's functions can be found from the spectral functions of each species:
$$F_K(\omega) = A_F(\omega) \coth\frac{\beta\omega}2\,, \quad G_K(\omega) = A_G(\omega)\tanh\frac{\beta\omega}{2} \,.$$
With this, we find that the imaginary part of the self-energy, i.e., the quasi-particle decay rate, can be written as:
\begin{equation}
    \Im\{\Sigma_R(k,\Omega)\} = \frac{J^2 \xi^2}{4\pi} \int \frac{d\omega}{2\pi} \int \frac{d^2q}{(2\pi)^2}  \int \frac{d^2q'}{(2\pi)^2} \cos^2 2\theta_{k-q'/2} A_F(q,\omega) A_G(-k+q',\omega-\Omega) \left(\coth \frac{\beta\omega}{2} - \tanh\frac{\beta(\omega-\Omega)}{2} \right)
    \label{eq:appSigmaIm}
\end{equation}

The momentum integrals over the boson momentum $q$ and the momentum carried by the disorder $q'$ are independent and can be performed separately. Integrating over the $q$ dependent terms of \eqref{eq:appSigmaIm} gives,
\begin{equation*}
\begin{split}
    & \int^{\pi/\xi}_0 \frac{d^2q}{(2\pi)^2}A_F(q,\omega) = -2\int^{\pi/\xi}_0 \frac{qdq}{2\pi} \Im\left\{ \big(\Delta_b + q^2/2M_b - \eta\omega -\im\alpha\omega \big)^{-1} \right\} \\
    &\quad \quad = -\frac{2M_b}{\pi} \left(\pi\Theta(\eta\omega-\Delta_b) + \tan^{-1} \frac{\alpha \omega}{\Delta_b - \eta\omega} \right) \\
    & \quad \quad \simeq -\frac{2M_b}{\pi} \left(\tan^{-1} \frac{\alpha \omega}{\Delta_b} \right)
\end{split}
\end{equation*}
Here in the third line we have made use of the fact that $\eta$ is an order of magnitude smaller than $\alpha$. 

On the other hand, integrating over the $q'$ dependent terms of \eqref{eq:appSigmaIm} gives,
\begin{equation*}
\begin{split}
    & \int^{\pi/\xi}_0 \frac{d^2q'}{(2\pi^2)} \cos^2 2\theta_{k-q'/2} A_G(-k+q'/2,\omega) = 2 \int \frac{dq'_{\parallel}}{2\pi} \frac{dq'_{\bot}}{2\pi} \cos^2 (2\theta_k - q'_{\parallel}/k_F) \Im\left\{\big(\omega-\Sigma_R(k,\omega)-v_F \Delta k + v_F q_{\bot} \big)^{-1}\right\} \\
    & \quad \quad = \frac{1}{2\pi v_F} \int_{-\pi/\xi}^{\pi/\xi} dq_{\parallel} \cos^2 (2\theta_k - q_{\parallel}/k_F)
\end{split}
\end{equation*}
Here, we have made use of the assumption that at low temperatures, our effective theory is limited to momentum $k$ is close to the fermi surface, so that $|k-k_F| \ll \frac{\pi}{\xi}$. $q'_{\bot}$ ($q'_{\parallel}$) denotes the momentum perpendicular (parallel) to the fermi surface, respectively.

Therefore, combining these two results, we find that the imaginary part of the self-energy is given as,
\begin{equation}
\begin{split}
    &\Im\{\Sigma_R(k,\omega)\}= -\gamma_k \int \frac{d\omega}{2\pi} \tan^{-1} \frac{\alpha \omega}{\Delta_b} \left(\coth \frac{\beta\omega}{2} + \tanh\frac{\beta(\Omega-\omega)}{2} \right) \\
    %& \simeq -\gamma_k \Bigg[ 2\int \frac{d\omega}{2\pi} \tan^{-1} \frac{\alpha \omega}{\Delta_b}\sinh^{-1}\beta\omega + \int \frac{d\omega}{2\pi} \left( \tan^{-1} \frac{\alpha \omega}{\Delta_b} - \tan^{-1}\frac{\alpha (\omega+\Omega)}{\Delta_b} \right) \tanh\frac{\beta\omega}{2} \Bigg] \\
    & \simeq  -\gamma_k T \log \frac{\alpha T}{\Delta_b} - \gamma_k T  f(\Omega,\Delta_b) \,, \textrm{ where }
    f(\Omega,\Delta_b) = \int \frac{d\omega}{2\pi} \left( \tan^{-1} \frac{\alpha \omega}{\Delta_b} - \tan^{-1}\frac{\alpha (\omega+\Omega)}{\Delta_b} \right) \tanh\frac{\beta\omega}{2}
    \end{split}
\end{equation}
Let us make a few remarks about this dimensionless function $f$. By definition, $f(0,\Delta_b) = 0$.
Also, at low frequencies,
$$f(\Omega, \Delta_b) \simeq \frac{\pi + \pi\tan^2\Delta_b/2\alpha T}{8} (\beta\Omega)^2 $$
On the other hand, for large frequencies $\Omega \gg \Delta_b$,  $f(\Omega,\Delta_b) \simeq \pi \log \cosh \frac{\beta\Omega}{2}$.

\section{Quantum Boltzmann Equation}\label{app:QBE}
The quantum Boltzmann equations model the evolution of a generalized probability distribution and allows for the calculation of transport even in the absence of sharply defined quasiparticles, such as in our model.
The quantum Boltzmann equation in the DC limit reads \cite{mah00},
\begin{subequations}
\label{eq:appBE}
\begin{align}
    & A_F(q,\Omega)^2 \Im\{\Pi_R(q,\Omega)\} \frac{2e q \cdot E}{M_b} {b_0}'(\Omega) = \Pi^> F^< - \Pi^< F^> \,,
    \label{eq:appBEb} \\
    & A_G(k,\omega)^2 \Im\{\Sigma_R(k,\omega)\} \frac{e k \cdot E}{m_f} {f_0}'(\omega) = \Sigma^> G^< - \Sigma^< G^> \,.
    \label{eq:appBEf}
\end{align}
\end{subequations}
Here, $f_0, b_0 = \frac{1}{e^{\beta\omega} \pm 1}$ denote the Fermi-Dirac/Bose-Einstein distribution functions, and $A_{G,F}$, the fermion/boson spectral functions.
The left hand side denotes the evolution of the boson and fermion distribution function in response to the external field, and the right hand side, the collision integral that account for the relaxation processes due to the fermion - boson scattering.

In the presence of an external field, the greater and lesser Green's functions shift from their equilibrium value and we have,
\begin{equation}
    G^{<,>}(k,\omega) = \frac{A_G(k,\omega)}{e^{\pm\beta\omega} + 1} + \delta f(k,\omega) \,, \quad
    F^{<,>}(q,\Omega) = \frac{A_F(q,\Omega)}{e^{\pm\beta\Omega} - 1} + \delta b(q,\Omega) \,,
    \label{eq:appGF<>}
\end{equation}
where $\delta f$ and $\delta b$ denote stand for the deviation from the fermion and boson equilibrium distribution.
As explained in the main text, the fermion and boson currents are then determined from $\delta f$ and $\delta b$, from which we subsequently find the charge conductance.

In this appendix, we derive the simplified quantum Boltzmann equations \eqref{eq:BE} which was used in the main text to determine $\delta f$ and $\delta b$.
We start with the derivation of the boson quantum Boltzmann equation \eqref{eq:BEb} from \eqref{eq:appBEb}. To this end, we require the greater and lesser boson self-energies $\Pi^{>,<}$ which appear in the collision integral.
These are given as,
$$\Pi^{>,<}(q,\Omega) = \int_{|k+k'-q|<\frac\pi\xi} \frac{d^2 k d^2k' d\omega}{(2\pi)^5} 2J^2 \xi^2 \cos^2 2\theta_{\hat{k}-\hat{k'}} G^{>,<}(k,\omega+\Omega) G^{>,<}(k',-\omega) \,.$$
At equilibrium, $G^{>,<}(k,\omega) = \frac{A_G(k,\omega)}{e^{\pm\beta\omega}+1}$ and we find,
\begin{equation}
\begin{split}
    \Pi^>(q,\Omega) & \simeq \frac{J^2 k_F \xi}{2\pi v_F^2} \frac{\Omega}{1-e^{-\beta\Omega}} \,, \ \
    \Pi^<(q,\Omega) \simeq \frac{J^2 k_F \xi}{2\pi v_F^2} \frac{\Omega}{e^{\beta\Omega}-1}
    \label{eq:appPi><}
\end{split}
\end{equation}
%This is in good accordance with standard expectations:
%When at equilibrium, the greater and lesser boson Green's functions follow the relation, $F^{<(>)}(q,\Omega) = A_F(q,\Omega) / (e^{\pm \beta\Omega}-1)$.
%Applying this with \eqref{eq:appPi><} to the collision integral, the right hand side of \eqref{eq:appBEb} vanishes as it should.
%In the presence of an electric field, we expect the greater and lesser boson Green's function to deviate a small amount from the equilibrium value.
%Let us denote this deviation as $\delta b(q,\Omega)$, so that $F^{<(>)}(q,\Omega) = A_F(q,\Omega) / (e^{\pm \beta\Omega}-1) + \delta b(q,\Omega)$. Applying this to \eqref{eq:appBEb} and dividing either side by $\alpha\Omega$ brings us to \eqref{eq:BEb} of the main text.
Applying \eqref{eq:appGF<>} and \eqref{eq:appPi><} together on the collision integral, and dividing either side by $\alpha\Omega$ brings us to \eqref{eq:BEb} of the main text.

We now derive the fermion quantum Boltzmann equation \eqref{eq:BEf} from \eqref{eq:appBEf}.
Foremost, the non-equilibrium fermion self-energy is given by,
\begin{equation}
\begin{split}
    \Sigma^>(k,\omega) &= \int \frac{d^2q d^2q' d\omega'}{(2\pi)^5} 2 J^2 \xi^2 \cos^2 2\theta_{k-\frac{q+q'}{2}} F^>(q,\omega+\omega') G^<(-k+q+q',\omega') \\
    &= \int \frac{d^2q d^2q' d\omega'}{(2\pi)^5} 2 J^2 \xi^2 \cos^2 2\theta_{k-\frac{q+q'}{2}} \big( b_0(\omega+\omega') + 1 \big) A_F(q,\omega+\omega') G^<(-k+q+q',\omega') \\
    %&= \int_{q,q',\omega'} J^2 \xi^2 \cos^2 2\theta_{k-\frac{q+q'}{2}} \big( b_0(\omega+\omega') + 1 \big) \frac{\alpha (\omega+\omega')}{\left(\Delta_b + \frac{q^2}{2 M_b} - \eta (\omega+\omega') \right)^2 + \alpha^2 (\omega+\omega')^2} G^<(-k+q+q',\omega') \\
    & \simeq \int \frac{d^2 q' d\omega'}{(2\pi)^3} \frac{2J^2 M_b \xi^2}{\pi} \cos^2 2\theta_{k-\frac{q+q'}{2}} \big( b_0(\omega+\omega') + 1 \big) \tan^{-1} \frac{\alpha(\omega+\omega')}{\Delta_b} G^<(-k+q+q',\omega') \,.
\end{split}
\end{equation}
In the second line, we have made the assumption that the bosons are in thermal equilibrium. 
This amounts to neglecting ``drag'' effects, by which the boson fluid is driven out of equilibrium when the fermions carry a current. Because of the disordered nature of the interaction, fermion-boson scattering does not conserve momentum, and hence drag effects are not likely to qualitatively change our results.
% This is justified since the typical momentum of bosons being $q \ll 1/\xi$, they rapidly reach thermal equilibrium through the scattering processes that preserve momentum only up to order $1/\xi$. In the third line, we have integrated over the boson momentum.

We now perform a change of variables from $k = (k_F + \Delta k) \hat{k}$ to $\hat{k}$ and $\Delta k = |k|-k_F$. The self-energy is mostly independent of $\Delta k$ allowing us to write $\Sigma_R(k,\omega) = \Sigma_R(\hat{k},\omega)$.
We now define $f(\hat{k},\omega)$ as the generalized distribution function for fermions pointing in the $\hat{k}$ direction with frequency $\Omega$. They are given as,
$$f(\hat{k},\omega) = v_F \int_{\Delta k} G^< \big((k_F + \Delta k)\hat{k},\omega \big) \,, \ \
1-f(\hat{k},\omega) = v_F \int_{\Delta k} G^> \big((k_F + \Delta k)\hat{k},\omega \big)$$

Upon this change of variables (and integrating out $\Delta k$), \eqref{eq:appBEf} becomes,
\begin{equation}
\begin{split}
    e \hat{k} \cdot E f_0'(\omega) &= \frac{16\pi}{v_F} \int_{|\hat{k} + \hat{k'}| < \frac\pi{k_F\xi}} \frac{d\hat{k'} d\omega'}{(2\pi)^2} \cos^2 2\theta_{\hat{k} - \hat{k'}} \tan^{-1} \frac{\alpha(\omega+\omega')}{\Delta_b} \\
    & \quad \Big[ \big(b_0(\omega+\omega') + 1 \big)  f(\hat{k},\omega) f(\hat{k'},\omega') - b_0(\omega+\omega') \big(1 - f(\hat{k},\omega) \big) \big(1 - f(\hat{k'},\omega') \big) \Big] \\
    &= \frac{16\pi}{v_F} \int_{|\hat{k} + \hat{k'}| < \frac\pi{k_F\xi}} \frac{d\hat{k'} d\omega'}{(2\pi)^2} \cos^2 2\theta_{\hat{k} - \hat{k'}}  B(\omega+\omega') (1 - f_0(\omega))(1 - f_0(\omega')) \big\{g(\hat{k},\omega) + g(\hat{k'},\omega') \big\} \\
    & \textrm{ where } B(\nu) = \tan^{-1} \frac{\alpha \nu}{\Delta_b} \ b_0(\nu)
    \label{eq:appBEf_mod}
\end{split}
\end{equation}
where $f(\hat{k},\omega) = f_0(\omega) + \delta f(\hat{k},\omega)$, and $\delta f(\hat{k},\omega) = f_0(\omega) (1-f_0(\omega)) g(\hat{k},\omega)$. This ultimately brings us to \eqref{eq:BEf} of the main text.

\subsection{Potential Disorder}
We now consider the effects of the following potential disorder acting on the fermions.
\begin{equation*}
    H_{rp} = V_{x} c^\dagger_x c_x\,, \textrm{ where } \overline{V_x} = 0, \ \overline{V_{x}V_{x'}} \sim W\delta_{x x'}\,.
\end{equation*}
This random potential results in an additional self-energy $\Sigma^{>,<}_{rp}(\omega) = W \int_{k} G^{>,<}(k,\omega)$ on the fermions.
In turn, this extra self energy results in an additional term in the collision integral of the fermion quantum Boltzmann equation and we have,
\begin{equation}
\begin{split}
    e \hat{k} \cdot E f_0'(\omega) &= \frac{16\pi}{v_F} \int_{|\hat{k} + \hat{k'}| < \frac\pi{k_F\xi}} \frac{d\hat{k'} d\omega'}{(2\pi)^2} \cos^2 2\theta_{\hat{k} - \hat{k'}}  B(\omega+\omega') \big(1 - f_0(\omega) \big) \big(1 - f_0(\omega') \big) \big\{g(\hat{k},\omega) + g(\hat{k'},\omega') \big\} \\
    &\quad + \frac{1}{\tau_0 v_F} f_0(\omega) \big(1 - f_0(\omega) \big) g(\hat{k},\omega)
    \label{eq:appBEf_dd}
\end{split}
\end{equation}

This means that \eqref{eq:BEf_rescaled}, the equation for $\tilde{g}$ is modified as,
\begin{equation}
\begin{split}
    \frac{\cos \theta_{\hat{k}} \tilde f_0(\tilde \omega) }{16\pi} &= \int_{|\hat{k} + \hat{k'}| < \frac\pi{k_F\xi}} \frac{d\hat{k'} d\tilde\omega'}{(2\pi)^2} \cos^2 2\theta_{\hat{k} - \hat{k'}}  \tilde B(\tilde\omega+\tilde\omega') \big(1 - \tilde f_0(\tilde\omega') \big) \big\{ \tilde g(\hat{k},\tilde\omega) + \tilde g(\hat{k'},\tilde\omega') \big\} + \frac{\tilde{f_0}(\tilde\omega)}{16\pi \tau_0 T}\tilde{g}(\hat{k},\tilde\omega) \,.
    \label{eq:appBEf_dd_rescaled}
\end{split}
\end{equation}

Solving \eqref{eq:appBEf_dd_rescaled} for $\tilde g$, and applying it to \eqref{eq:sigmaf_d} we can obtain the longitudinal conductivity as a function of $\tau T$ and $\xi$.

\subsection{Magnetotransport}
We now turn to magnetotransport within our model.
We will derive the quantum Boltzmann equation used in the main text \eqref{eq:BEf_rescaled_B} to describe transport under a magnetic field.
Furthermore, we will demonstrate that the boson magnetotransport is negligble in comparison to that of the fermions.

The QBE \eqref{eq:BE} is modified in the following manner in the presence of a perpendicular magnetic field $B$:
\begin{subequations}
\label{eq:appBEB}
\begin{align}
    & A_F(q,\Omega)^2 \Im\{\Pi_R(q,\Omega)\} \frac{2e q \cdot E}{M_b} {b_0}'(\Omega) + \frac{2e q \times B \hat{z}}{M_b} \cdot \nabla_q F^< = \Pi^> F^< - \Pi^< F^> \,,
    \label{eq:appBEb_B} \\
    & A_G(k,\omega)^2 \Im\{\Sigma_R(k,\omega)\} \frac{e k \cdot E}{m_f} {f_0}'(\omega) + \frac{e k \times B \hat{z}}{m_f} \cdot \nabla_k G^< = \Sigma^> G^< - \Sigma^< G^> \,.
    \label{eq:appBEf_B}
\end{align}
\end{subequations}

We first derive the fermion quantum Boltzmann equation of the main text \eqref{eq:BEf_rescaled_B}.
As in the derivation of \eqref{eq:appBEf_mod}, we integrate over $\Delta k$:
We find that \eqref{eq:appBEf_B} then becomes,
\begin{equation}
\begin{split}
    &e \hat{k} \cdot E f_0'(\omega) - \frac{\omega_{cf}}{v_F} f_0(\omega) \big(1 - f_0(\omega) \big) \partial_{\theta_k} g(\hat{k},\omega) \\
    & \qquad =  \frac{16\pi}{v_F} \int_{|\hat{k} + \hat{k'}| < \frac\pi{k_F\xi}} \frac{d\hat{k'} d\tilde\omega'}{(2\pi)^2} \cos^2 2\theta_{\hat{k} - \hat{k'}}  B(\omega+\omega') \big(1 - f_0(\omega) \big) \big(1 - f_0(\omega') \big) \big\{g(\hat{k},\omega) + g(\hat{k'},\omega') \big\} \,,    \label{eq:appBEf_b_mod}
\end{split}
\end{equation}
where $\omega_{cf} = \frac{eB}{m_f}$ denotes the fermion cyclotron frequency.
Rescaling frequency with temperature, we arrive at \eqref{eq:BEf_rescaled_B} of the main text and find that the magnetic field influences the fermion conductance as a function of $b = \frac{\omega_{cf}}{T}$.

We now turn to boson magnetotransport. There, the rotation symmetry facilitates a simple solution to \eqref{eq:appBEb_B}.
Let us switch to polar coordinates in momentum space. We find that \eqref{eq:appBEb_B} is given as,
\begin{equation*}
    \alpha \Omega A_F(|q|,\Omega)^2 \frac{2e |q| E \cos\theta_q}{M_b} {b_0}'(\Omega) + \omega_{cb} \nabla_{\theta_q} \delta b(|q|,\theta_q,\Omega) = \alpha \Omega \delta b(|q|,\theta_q,\Omega) \,.
\end{equation*}
Here, $\omega_{cb} = \frac{2e B}{M_b}$ denotes the boson cyclotron frequency, and we have set the electric field in the $\hat{x}$ direction without loss of generality.
Performing a Fourier transform in the angular coordinate $\theta_q$, we find that \eqref{eq:appBEb_B} simplifies to,
\begin{equation}
    A_F(|q|,\Omega)^2 \frac{2e|q|E}{M_b} b_0'(\Omega) \frac12 \{\delta_{l,1} + \delta_{l,-1}\} = \left(1 - \frac{\im l \omega_{cb}}{\alpha\Omega} \right) \delta b_l(|q|,\Omega)
    \label{eq:appBEb_B_ang}
\end{equation}
where $\delta b_l(|q|,\Omega) = \int_{\theta_q} e^{-\im l \theta_q} \delta b(|q|,\theta_q,\Omega)$ denotes the nonequilibrium boson distribution with angular momentum $l$.

From \eqref{eq:appBEb_B_ang}, we find that $\delta b_l$ is zero other than for $l = \pm 1$, which are given by,
\begin{equation*}
    \delta b_{\pm 1}(q,\Omega) = \frac{eq E b_0'(\Omega)}{M_b} \frac{A_F(q,\Omega)^2}{1 \mp \frac{\im \omega_{cb}}{\alpha\Omega}}
\end{equation*}
The longitudinal and Hall currents can be calculated from $\delta b_{1,-1}$ through the relation,
\begin{equation*}
\begin{split}
    & J_{x} = \int \frac{d^2 q d\Omega}{(2\pi)^3} \frac{2 e q \cos \theta}{M_b} \delta b(q,\Omega) = \int \frac{d|q| d\Omega}{(2\pi)^2} \frac{e q^2}{2\pi M_b} \big( \delta b_{1}(q,\Omega) + \delta b_{-1}(q,\Omega) \big) \,, \\ 
    & J_{y} = \int \frac{d|q| d\Omega}{(2\pi)^2} \frac{\im e q^2}{2\pi M_b} \big( \delta b_{1}(q,\Omega) - \delta b_{-1}(q,\Omega) \big) \,.
\end{split}
\end{equation*}
Putting this altogether, we find that the boson magnetotransport can be computed through the following integral equations:
\begin{equation}
\begin{split}
    \sigma_b^{xx} &= \int \frac{d|q| d\Omega}{(2\pi)^2} \frac{e^2 q^3 b_0'(\Omega) A_F(q,\Omega)^2}{2\pi M_b^2} \frac{\alpha^2 \Omega^2}{\alpha^2 \Omega^2 + \omega_{cb}^2} \,, \\
    &= \frac{e^2}{4\pi^3} \int_{-\infty}^{\infty} d\tilde\Omega \int_{0}^\infty d \tilde x \frac{\tilde x \tilde \Omega^2 \csch^2\frac{\tilde\Omega}{2}}{\big(\tilde\Delta_b / \alpha + \tilde x - \frac{\eta}{\alpha} \tilde\Omega \big)^2 + \tilde \Omega^2} \frac{\tilde\Omega^2}{\tilde\Omega^2 + \tilde b^2} \\
    \sigma_b^{xy} &= \int \frac{d|q| d\Omega}{(2\pi)^2} \frac{e^2 q^3 b_0'(\Omega) A_F(q,\Omega)^2}{2\pi M_b^2} \frac{\alpha \Omega \omega_{cb}}{\alpha^2 \Omega^2 + \omega_{cb}^2} \\
    &= \frac{e^2}{4\pi^3} \int_{-\infty}^{\infty} d\tilde\Omega \int_{0}^\infty d \tilde x \frac{\tilde x \tilde \Omega^2 \csch^2\frac{\tilde\Omega}{2}}{\big(\tilde\Delta_b / \alpha + \tilde x - \frac{\eta}{\alpha} \tilde\Omega \big)^2 + \tilde \Omega^2} \frac{\tilde\Omega \tilde b}{\tilde\Omega^2 + \tilde b^2}
    \label{eq:appsigmabB}
\end{split}
\end{equation}

We see from \eqref{eq:appsigmabB} that the boson conductance is a function of $\tilde \Delta_b/\alpha$, $\eta/\alpha$, and $\tilde b = \frac{\omega_{cb}}{\alpha T}$. Here, $\tilde \Delta_b/\alpha$ and $\eta/\alpha \sim (k_F\xi)^{-1}$ are constants independent of temperature.
Therefore, the dimensionless ratio $\tilde b \sim (k_F\xi) \frac{\omega_{cf}}{T}$ determines the boson magnetotransport.

\begin{figure}
    \centering
    \includegraphics[width = 0.3 \columnwidth]{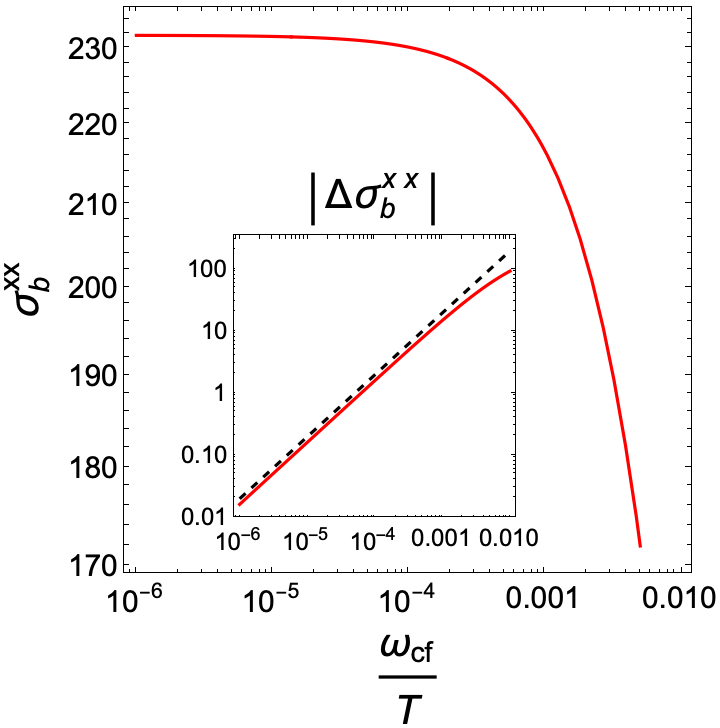}
    \includegraphics[width = 0.31 \columnwidth]{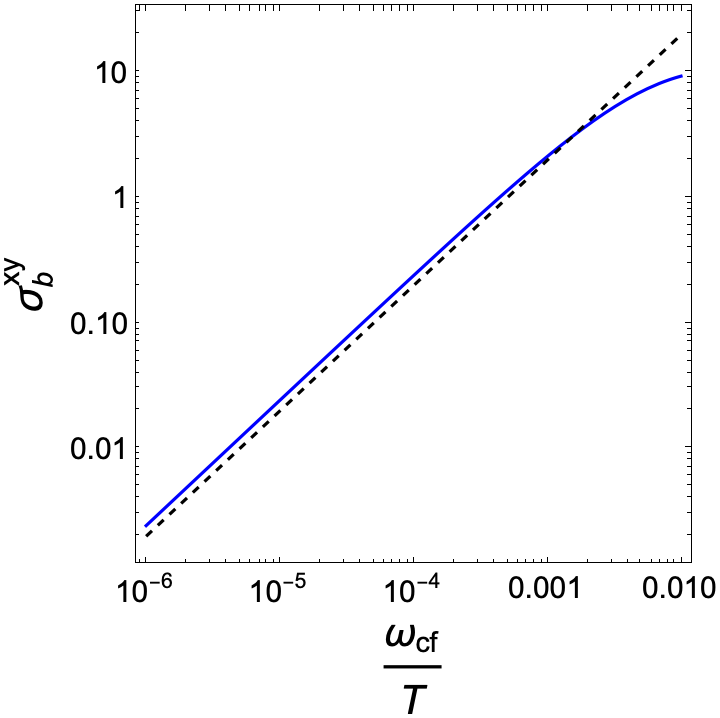}
    \caption{$\sigma_b^{xx}, \ \sigma_b^{xy}$ as a function of $b = \omega_{cf}/T$, plotted in log-log scale for $\tilde \Delta_b = 1$, $\alpha = 10$, $\eta = 1$ which roughly corresponds to $k_F\xi \sim 10$. Inset of the $\sigma_b^{xx}$ plot shows $|\Delta\sigma_b^{xx}| = \sigma_b^{xx}(0,T) - \sigma_b^{xx}(B,T)$ as a function of $b$ in log-log scale. Dashed lines in both the left inset and right figures are of slope 1.}
    \label{fig:appsigmab}
\end{figure}

In Fig.\ref{fig:appsigmab} we display the magnetic field dependence of $\sigma_b^{xx,xy}$ obtained by numerically integrating \eqref{eq:appsigmabB}. For direct comparison with fermion magnetotransport, we plot the x axis as a function of $b = \frac{\omega_{cf}}{T}$.
We see that the change in the longitudinal boson conductivity $|\Delta \sigma_b^{xx}|$ scales linearly with $\frac{\omega_{cf}}{T}$.
At low temperatures, this change is minuscule in comparison to that of the fermions.
In the linear magnetoresistance regime of $b \in [(k_F\xi)^{-4}, (k_F\xi)^{-3} ]$, recall that $\sigma_f^{xx} \sim \epsilon_F/\omega_{cf} = \epsilon_F/T b^{-1}$.
For typical temperatures $\epsilon_F/T$ is a large number, and so is $b^{-1}$ in the LMR regime. %ranging from $[10^2, 10^4]$ for typical temperatures ranging from $1K$ to $100K$
Therefore, the change in the boson magnetoresistance that scales linearly with $b$ is completely negligible.

\section{Diamagnetic Susceptibility}\label{app:dia}
In order to compute the diamagnetic susceptibility, we take the current-current correlator $\braket{J \cdot J}$ and set $\Omega = 0$ and take the limit of $q \rightarrow 0$.  The diamagnetic susceptibility is given as,
\begin{equation*}
    \chi = -\lim_{q \rightarrow 0} \partial_{q}^2 \braket{J_x \cdot J_x} (q \hat{y},\omega=0)
\end{equation*}

Let us for simplicity assume a quadratic dispersion for the fermions, $\epsilon_k = \frac{k^2}{2m_f}$. Then we find,
\begin{equation}
\begin{split}
    \chi_f &= \frac{2e^2}{(2\pi)^2 \beta m_f^2} \partial_{q}^2 \sum_{\omega_n} \int \frac{d^2 k}{(2\pi)^2} k^2 \cos^2{\theta} \ G_{k+q\hat{y}/2}(\omega_n) G_{k-q\hat{y}/2}(\omega_n)  \\
    &= -\frac{2e^2}{(2\pi)^2 \beta m_f^2} \sum_{\omega_n} \int \frac{d^2 k}{(2\pi)^2} \frac{ \frac{k^2}{2m_f} \cos^2{\theta} \Big( \frac{k^2}{2m_f} (\cos^2 \theta - \sin^2 \theta) - \mu + \im\omega_n + \Sigma(k,\omega_n) \Big) }{\Big(\frac{k^2}{2m_f} - \mu + \im\omega_n + \Sigma(k,\omega_n) \Big)^4} \\
    %& = -\frac{e^2}{\pi \beta m_f^2} \sum_{\omega_n} \int_k k^3 \frac{\mu-\Sigma(k,\omega)}{4m_f\big(-\im\omega_n - \Sigma(k,\omega_n) - (\epsilon_k-\mu) \big)^4} \\
    & = -\frac{e^2}{4\pi \beta m_f} \sum_{\omega_n} \int \frac{d\epsilon_k}{2\pi} \frac{\epsilon_k^2 + 2 \epsilon_k \big(-\im\omega-\mu-\Sigma(k,\omega) \big)}{\big(-\im\omega_n - \Sigma(k,\omega_n) - (\epsilon_k-\mu) \big)^4} \\
    &= -\frac{e^2}{12\pi m_f \beta} \sum_{\omega_n} \frac{1}{\im\omega_n+\Sigma-\mu} \\
    &\simeq \frac{e^2}{12\pi m_f}
\end{split}
\end{equation}
In the third line we used the fact that the self-energy $\Sigma \in T[(k_F \xi)^{-3}, (k_F \xi)^{-1}]$ and is small. Consequently, we find that the low temperature diamagnetic response of fermions $\chi_f$ is similar to the Landau diamagnetism of free fermions. Similar results have been seen in \cite{Aldape:2020enq}.

We now calculate the diamagnetic response due to Cooper pairs. Evaluating the boson current-current correlator, we find,
\begin{equation}
\begin{split}
    \chi_b &= -\frac{4e^2}{(2\pi)^2 \beta M_b^2} \partial_{q}^2 \sum_{\omega_n} \int \frac{d^2 k}{(2\pi)^2} k^2 \cos^2{\theta} \ F_{k+q\hat{y}/2}(\omega_n) F_{k-q\hat{y}/2}(\omega_n) \\
    %\frac{4e^2}{(2\pi)^2\beta q^2} \sum_{\omega_n} \int_k \frac{k^2}{M_b^2} [F(k+q/2,\omega_n) F(k-q/2,\omega_n) - F(k,\omega_n)^2] \\
    %&= \frac{4e^2}{(2\pi)^2\beta q^2 M_b^2} \sum_{\omega_n} \int_k \frac{k^2}{\left(\Delta_b + k^2/2M_b + q^2 \cos^2\theta/8M_b + \alpha|\omega_n|\right)^2 - k^2q^2/4M_b^2} - \frac{k^2}{(\Delta_b+k^2/2M_b+\alpha|\omega_n|)^2} \\
    %&= \frac{e^2}{8\pi^2 \beta} \sum_{\omega_n} \int_k \frac{k^2/M_b^2(\Delta_b+\alpha|\omega_n|)}{(\Delta_b+k^2/2M_b+\alpha|\omega_n|)^4} \\
    &= \frac{e^2}{6\pi\beta M_b}\sum_{\omega_n}\frac{1}{\Delta_b
    +\alpha|\omega_n|} \\
\end{split}
\end{equation}
which brings us to $\chi_b$ in \eqref{eq:chib}.

\end{widetext}
\end{document}